\documentclass[longbibliography,twocolumn,superscriptaddress,showpacs]{revtex4-1}

\usepackage{CJK}

\usepackage{graphicx}

\usepackage{amsmath}
\usepackage{dcolumn}

\usepackage{bm}
\usepackage[normalem]{ulem}
\usepackage{color}
\usepackage{hyperref}

\usepackage{epstopdf}
\makeatletter
\def\maketitle{
	\@author@finish
	\title@column\titleblock@produce
	\suppressfloats[t]}
\makeatother

\newcommand{\bew}{\begin{widetext}}
\newcommand{\ew}{\end{widetext}}

\newcommand{\ii}{{\rm i}}

\newcommand{\bp}{\mathbf{p}}
\newcommand{\bq}{\mathbf{q}}
\newcommand{\bv}{\mathbf{v}}

\newcommand{\br}{\mathbf{r}}

\newcommand{\bff}{\mathbf{f}}
\newcommand{\bu}{\mathbf{u}}

\newcommand{\bbr}{\mathbf{r}}

\newcommand{\bk}{\mathbf{k}}

\newcommand{\bh}{\mathbf{h}}

\newcommand{\beq}{\begin{equation}}
\newcommand{\eeq}{\end{equation}}
\newcommand{\beqn}{\begin{eqnarray}}
\newcommand{\eeqn}{\end{eqnarray}}
\newcommand{\pp}{\partial}
\newcommand{\dd}{{\rm d}}

\newcommand{\cG}{{\cal G}}

\newcommand{\la}{\langle}

\newcommand{\ra}{\rangle}

\newcommand{\vnab}{{\bf \nabla}}

\begin{document}
\title{Dynamics of packed swarms: time-displaced correlators of two dimensional incompressible flocks}
\author{Leiming Chen}
\email{leiming@cumt.edu.cn}
\address{School of Material Science and Physics, China University of Mining and Technology, Xuzhou Jiangsu, 221116, P. R. China}
\author{Chiu Fan Lee}
\email{c.lee@imperial.ac.uk}
\address{Department of Bioengineering, Imperial College London, South Kensington Campus, London SW7 2AZ, U.K.}
\author{Ananyo Maitra}
\email{nyomaitra07@gmail.com}
\address{Laboratoire de Physique Th\'eorique et Mod\'elisation, CNRS UMR 8089,
	CY Cergy Paris Universit\'e, F-95032 Cergy-Pontoise Cedex, France}
\address{{Sorbonne Universit\'{e} and CNRS, Laboratoire Jean Perrin, F-75005, Paris, France}}
\author{John Toner}
\email{jjt@uoregon.edu}
\affiliation{Department of Physics and Institute of Theoretical
Science, University of Oregon, Eugene, OR $97403^1$}
\affiliation{Max Planck Institute for the Physics of Complex Systems, N\"othnitzer Str. 38, 01187 Dresden, Germany}

\begin{abstract}
{We analytically calculate the scaling exponents of a two-dimensional KPZ-like system: coherently moving incompressible polar active fluids. Using three different renormalization group approximation schemes, we obtain values for the  ``roughness" exponent $\chi$ and  anisotropy exponent $\zeta$ that are extremely near the known exact results. This implies our prediction for the previously completely unknown dynamic exponent $z$ is quantitatively accurate.}
	\end{abstract}
	
\maketitle
The Mermin-Wagner theorem \cite{MW} states that in equilibrium, two-dimensional systems with finite-range interactions cannot spontaneously break a continuous symmetry at any finite temperature.
One of the earliest and most surprising results in the field of active matter was that there is no analogue of the Mermin-Wagner theorem in nonequilibrium matter. In particular, it was \emph{explicitly} demonstrated that active polar units moving on a two-dimensional frictional substrate and with
\emph{purely} short-range interactions \emph{can} spontaneously break continuous rotation symmetry in two dimensions and form long-range ordered flocks \cite{vicsek_prl95, toner_prl95, toner_pre98, toner_pre12, Chate_DADAM_rev}. Such flocks have a \emph{non-zero} mean speed $\langle{\bf v}\rangle$ even at finite noise strengths.



Despite the fact that
the \emph{existence} of long-range-ordered two-dimensional flocks has been demonstrated analytically, determining their scaling behavior analytically has proved much more challenging.
One class of  systems that \emph{has} proved amenable to analytical treatment is  \emph{incompressible} flocks \cite{inc_trans, CLT_Ncomm, Chen, Ano_pol, CLMT-PRL1, CLMT-PRE1}.

Incompressibility can arise in
many ways. One way is to make the density extremely high. In this limit
the effective compressibility of the flockers vanishes, with any departure from the mean density being severely penalised \cite{CLT_Ncomm, Ano_pol}. More accurately, as the compressibility goes to $0$, the lengthscale up to which the system is effectively incompressible diverges. A flock formed by a suspension of active swimmers in an incompressible fluid in a narrow channel also inherits the incompressibility of the fluid and is therefore incompressible. This is technically distinct from the model we will consider here due to the presence of an extra conserved quantity -- the number of swimmers. However, if the number of active units is \emph{not} conserved, for instance, due to birth and death \cite{Toner_Malthus, CLT_malthus1, CLT_malthus2}, the dynamics of such a system does belong to the ``incompressible flock'' universality class we consider here.

Our understanding of the {\it equal-time} behavior of incompressible two dimensional flocks is quite complete. Although they are non-equilibrium systems, their hydrodynamic equations prove to be equivalent, once terms that are ``irrelevant" in the renormalization group sense are dropped,  to those of an equilibrium magnetic system \cite{CLT_Ncomm, Kashuba}  with long-ranged interactions. Because of those long-ranged interactions, the Mermin-Wagner theorem does not apply to these magnetic systems. The partition function for this equilibrium system can then be further mapped \cite{CLT_Ncomm} onto that for a two-dimensional smectic, whose equal-time scaling laws are known {\it exactly} via a further mapping \cite{Golub} onto the $1+1$-dimensional KPZ equation \cite{KPZ}. This analysis \cite{CLT_Ncomm} gives the scaling law for the equal time fluctuations   $\bu( \bbr,t)$ of the local active fluid velocity $ \bv(\bbr,t)$ about its mean value $\la \bv \ra\equiv v_0 \hat{\bf x}$, where we've defined our coordinate system so that ${\bf \hat{x}}$ is along the mean velocity spontaneously chosen by the system. Specifically 
\beq
\langle \bu(\br,t)\cdot\bu(\mathbf{0},t)\rangle=|x|^{2\chi}\cG_{_{ET}}\left({|y|\over |x|^{\zeta}}\right)\, ,\label{ETCorrel1}
\eeq
where ``ET" stands for  ``equal-time". 
The exponents $\chi$ and $\zeta$ were determined {\it exactly} by the aforementioned mappings\cite{CLT_Ncomm} to be  $\chi=-1/2$ and $\zeta=3/2$.

Unfortunately, since this analysis was based entirely on the partition function for the equivalent equilibrium model, no information about the {\it dynamics} of the system can be obtained by this exact mapping. Most of this missing information can be encoded in a single additional universal exponent, namely,  the ``dynamical" exponent $z$. This can be defined by considering  the {\it unequal}- time correlation function of the velocity fluctuations, which obeys the scaling law
\beqn
&&\langle \bu(\br,t)\cdot\bu(\mathbf{0},0)\rangle\nonumber
\\&=&|x -v_bt|^{2\chi}\cG_{_{UET}}\left({|y|\over |x-v_bt|^{\zeta}},{|t|\over |x -v_bt|^{z}}\right)\, ,\label{Correl1}
\eeqn
and the subscript ``UET" now means ``unequal-time". The "boost velocity $v_b$ is a phenomenological parameter of our model.

In this Letter, we obtain the dynamical exponent $z$ for the polar phase of two-dimensional, incompressible flocks using three different dynamic renormalization group (DRG)
schemes: two different uncontrolled calculations exactly in two dimensions,  and a $d=(d_c-\epsilon)$-expansion.
We get
\beq
z=1.65\pm0.06\,,\label{Dyna_Exp}
\eeq
where the error bars correspond to the averages over the three DRG schemes.
We also  calculate the the roughness exponent $\chi$ and the anisotropic exponent $\xi$ in all three schemes.

As mentioned earlier, we already know the static exponents $\chi$ and $\zeta$ exactly, so our purpose in performing this DRG is only to calculate $z$. However, our knowledge of the exact values of $\chi$ and $\zeta$ provides us with a useful check on the quantitative accuracy of our DRG calculations, because we can compare the approximate values of $\chi$ and $\zeta$ that we get from those schemes with the known exact values. And they prove to be very close; indeed, one of our uncontrolled approximations reproduces the known exact values
$\chi=-1/2$ and
$\zeta=3/2$.
We therefore believe our prediction (\ref{Dyna_Exp}) is very accurate quantitatively.

Inserting (\ref{Dyna_Exp}) and the exact value for $\chi$ into (\ref{Correl1}), and considering the limit $\br\to\mathbf{0}$, we obtain 
the temporal part of the velocity correlation:
\beq
\langle \bu(\mathbf{0},t)\cdot\bu(\mathbf{0},0)\rangle\sim At^{2\chi\over z}=At^{-0.61\pm 0.02}\, ,\label{Correl2}
\eeq
where $A$ is some non-universal constant.


We'll now present the drivation of the above results. This begins with
the hydrodynamic equation of motion (EOM) of
incompressible polar active fluids can be constructed purely based on symmetry considerations, as has been discussed in \cite{CLT_Ncomm}, and is:
\bew
\beq
\label{eq:maineom}
\pp_t \bv+ \lambda (\bv \cdot \vnab )\bv = -\vnab  \Pi  -(\bv \cdot \vnab \Pi_1) \bv +U(|\bv|) \bv
+\mu_1 \nabla^2 \bv +\mu_2 (\bv \cdot \vnab)^2 \bv
+\bff(\br,t)
\ ,
\eeq
\ew
where the ``pressure"   $\Pi$ is a Lagrange multiplier that enforces the incompressibility constraint: $\vnab \cdot \bv = 0$, the term involving  $U(|\bv|)$---a smooth, analytic function of $|\bv|$---ensures that there is an ``ordered'' phase in which $\bv$ has a nonzero mean magnitude $v_0$. That is, we assume that there is a regime in parameter space of $U(|\bv|)$ where $U(|\bv|)>0$ for $|\bv|<v_0$, $U(|\bv|)=0$ for $|\bv|=v_0$, and $U(|\bv|)<0$ for $|\bv|>v_0$. The ``anisotropic pressure" $\Pi_1$ is another generic function of $|\bv|$.
Finally, $\bff(\br)$ is a zero-mean, Gaussian white noise with the correlation
	\begin{equation}
		\langle f_i(\br,t)f_j(\br',t')\rangle=
		2D\delta_{ij}\delta^2(\br-\br')\delta(t-t')\, ,\label{Random_A}
	\end{equation}
where the indices $i,j$ enumerate the spatial coordinates.

Expanding \eqref{eq:maineom} about an ordered state using ${\bf v}=v_0\hat{{\bf x}}+{\bf u}$ and again retaining only relevant terms
 (i.e., the terms that are important in the limit of large time and length scales),
we obtain the EOM governing  $\bu$
\beqn
\label{eq:u_eom}
\pp_t u_i&=&-\pp_i \Pi+\mu_\perp \pp_y^2 u_i+\mu_x \pp_{x}^2 u_i
\nonumber\\
&&
-\alpha\left(u_x+{u_y^2\over 2v_0}\right)(\delta_{ix}+{u_y\over v_0}\delta_{iy})+f_i\,,
\eeqn
{where
$\alpha\equiv -v_0\left(\dd U\over\dd |\bv|\right)_{|\bv|=v_0}$,
$\mu_{_\perp}\equiv \mu_1$,
and $\mu_x\equiv\mu_1+\mu_2v_0^2$.
As discussed in \cite{CLT_Ncomm, Toner_Malthus},  we have performed a Galilean transformation to a reference frame ${\bf r}'$ which moves with respect to our original reference frame in the direction of mean flock motion at a speed $v_b=\lambda v_0$; that is, ${\bf r}'\equiv {\bf r}-\lambda v_0 t\hat{{\bf x}}$, or, equivalently, $x'=x-v_bt$.  For simplicity, in
(\ref{eq:u_eom}) we have dropped the prime in $x$.
Note that $u$ is also subject to the incompressibility constraint $\vnab \cdot \bu = 0$ inherited from $\vnab \cdot \bv = 0$. The somewhat non-trivial power-counting that leads to \eqref{eq:u_eom} is detailed in \cite{CLT_Ncomm}.

 We now perform a DRG analysis to obtain the dynamic exponent for the incompressible flock.
Fourier transforming \eqref{eq:u_eom} and acting on the projection operator  $P_{yi}(\bq)$ to eliminate the pressure term \cite{foot1}, we obtain
\begin{widetext}
	\beq
	-\ii\omega u_y=P_{yx}\left(\bq\right)\mathcal{F}_{\tilde{\bq}}
	\left[-\alpha\left(u_x+{u_y^2\over 2}\right)\right]+P_{yy}(\bq)\mathcal{F}_{\tilde{\bq}}
	\left[-\alpha
	\left(u_x+{u_y^2\over 2}\right)u_y+\mu_x\partial_x^2u_y+f_y\right]\,,
	\label{2Dy4}
	\eeq
\end{widetext}
where
we have eliminated $v_0$ by absorbing it into $\bu$ (i.e., $\bu\to v_0\bu$), and the symbol $\mathcal{F}_{\tilde{\bq}}$ represents the $\tilde{\bq}$th Fourier component.

Now we implement the standard DRG procedure \cite{forster_pra77} on EOM \eqref{2Dy4}. First we decompose the field $\bu$ into ``slow" and ``fast"  parts $u^<(\bq)$ and $u^>(\bq)$, which are supported at small $\bq$'s  and large $\bq$'s,
respectively.
(We will eventually take the large $\bq$'s to lie in an infinitesimally thin shell along the outer edge of the Brillouin zone).

 Next we average out $u^>(\bq)$ to get the effective EOM for $u^<(\bq)$. In this step the various coefficients in \eqref{2Dy4} are renormalized and their corrections can be represented by graphs (i.e., Feynman diagrams); these corrections are therefore called graphical corrections. Finally we rescale time, lengths, and  fields as
\begin{subequations}
\label{rescale}
\begin{align}
		t\to te^{z\dd\ell},~~x\to xe^{\dd\ell},~~y\to ye^{\zeta\dd\ell},\\
		u_y\to u_ye^{\chi\dd\ell},~~u_x\to u_xe^{\left(\chi+1-\zeta\right)\dd\ell}\,,
\end{align}
\end{subequations}
to restore  the Brillouin zone to its original size. Note that the form of the rescaling in $u_x$ is imposed by the incompressibility condition.

This  whole process can be repeated iteratively, which leads to recursion relations for the three coefficients $\alpha$, $\mu_x$, and $D$ in \eqref{2Dy4}  [Note that $D$ is hidden in the correlations of the noise $\bff$  (\ref{Random_A})].

In deriving these recursion relations, we use three distinct DRG  schemes. Two of these are one-loop order DRG calculations in exactly two dimensions, which are uncontrolled approximations since there is no limit in which they become exact.
The third scheme is an $\epsilon=5/2-d$ expansion to $O(\epsilon)$, in which
we analytically continue our calculation to $d>2$ by treating the $y$ direction as $d-1$ dimensional. This is a controlled approximation, since it formally becomes exact in the limit $\epsilon\to0$. Obviously,  it is also only approximate in $d=2$.
 The first of the two uncontrolled approximations  uses a Brillouin zone (BZ)
 $-\infty<q_y<\infty$, $-\Lambda<q_x<\Lambda$, where $\Lambda$ is the ultraviolet cut-off,  and the second uses the BZ
$-\infty<q_x<\infty$, $-\Lambda<q_y<\Lambda$. The $\epsilon$-expansion result is independent of the shape of the Brillouin zone. 
 All three approaches yield values of the dynamical exponent that are fairly close to each other.  Furthermore, all three obtain values of the already exactly known exponents $\chi$ and $\zeta$ which prove to be very close to those known exact values; indeed,  the first uncontrolled approximation we use yields the exact values of the
 $\chi$ and $\zeta$.

Crucially, in all three DRG calculations, we make use the symmetry properties of the EOM for ${\bf u}$ to make two important simplifications. First, we notice that rotation invariance of our hydrodynamic EOM is ensured by choosing the values of $\chi$ and $\zeta$ to keep all three ``$\alpha$"s   appearing in (\ref{2Dy4}) the same under rescaling, i.e.,
\beq
\chi=\zeta-1\,.\label{Chi1}
\eeq
This simplification reduces total number of the recursion relations to three. Specifically,
\begin{eqnarray}
	\frac{\dd\ln \alpha}{\dd \ell} &=& z-2 \zeta +2+\eta_\alpha\,,\label{Alpha_flow_1}\\
	\frac{\dd\ln \mu_x}{\dd \ell} &=& z-2+\eta_\mu\,,\label{Mu_flow_1}
\end{eqnarray}
and
\begin{equation}
	\frac{\dd\ln D}{\dd \ell} =z -3+\zeta+\eta_D \label{D_flow_un}
\end{equation}
for the uncontrolled approximations, or
\begin{equation}
	\frac{\dd\ln D }{\dd \ell} =z -3+(3-d)\zeta+\eta_D \label{D_flow_ep}
\end{equation}
for the $\epsilon$ expansion, where $\eta_{\alpha,\,u,D}$ denote graphical corrections.

Next, the fact that the dynamics of ${\bf u}$ obeys detailed balance introduces another simplification. This becomes clear if we formally introduce a friction coefficient in the dynamics for ${\bf u}$; $\Gamma\partial_t u_i=-\delta H/\delta u_i+f_i${, which is just the time-dependent-Ginzberg-Landau model, or model $A$ \cite{PCMP, Hohenberg}, with $H({\mathbf M})$ the Hamiltonian for a ordered divergence-free two-dimensional magnet expanded around its minimum at nonzero magnetization, where ${\mathbf M}=(v_0+u_x)\hat{x}+u_y\hat{y}$.} $\Gamma=1$ in \eqref{2Dy4} but it doesn't retain that value under renormalization. This would appear as a renormalization of the coefficient of $-i\omega u_y$; in {general}, this is an independent quantity. However, here detailed balance implies that the ratio $D/\Gamma$ \emph{cannot} change under renormalization i.e., $\Gamma$ and $D$ must renormalize in the same way. This implies that the correction to $\ln\Gamma$ are the same as the correction to $\ln D$. If we denote the direct graphical correction to the annealed noise $\eta^{dir}_D$ and the direct graphical correction to $\Gamma$ as $\eta_\omega$, the argument just given implies $\eta_\omega=\eta^{dir}_D$. To avoid retaining the friction coefficient as another extra parameter, we divide both sides of the renormalized EOM by the coefficient of $-i\omega u_y$ to {fix} the coefficient of $-i\omega u_y$ {at} $1$. This effectively introduces an additional correction $-\eta_D$ to both  $\alpha$ and $\mu$, and $-2\eta_D$ to $D$. (The factor of $2$ arises
because $D$ is proportional to the correlation of {\it two} noises).

Therefore, the overall graphical corrections to $\alpha$, $\mu$, and $D$ are given by}
\beqn
	&&\eta_\alpha=\eta^{dir}_\alpha-\eta^{dir}_D\,, ~~~\eta_\mu=\eta^{dir}_\mu-\eta^{dir}_D\,,\\
 &&\eta_D=\eta^{dir}_D-2\eta^{dir}_D=-\eta^{dir}_D\,.
\eeqn

Using  these considerations, and explicitly evaluating the graphical corrections to one-loop order, we find the following recursion relations 
for the first uncontrolled approximation, i.e., using the BZ  $-\infty<q_y<\infty$, $-\Lambda<q_x<\Lambda$ in exactly $d=2$:
\begin{eqnarray}
	\frac{\dd\ln \alpha}{\dd \ell}
	&=&z-2 \zeta +2-\frac{3g^{U1}}{4},\label{Alpha_flow_un_1}\\
	\frac{\dd\ln \mu_x}{\dd \ell}
	&=&z-2+\frac{g^{U1}}{4},\label{Mu_flow_un_1}\\
	\frac{\dd\ln D}{\dd \ell}
	&=&z -3+\zeta-\frac{g^{U1}}{4},\label{D_flow_un_1}
\end{eqnarray}
where
\begin{equation}
	g^{U1}=\frac{\alpha^{1/2}D}{4\mu^{3/2}\pi{\Lambda}}\,.\label{gu1def}
\end{equation}
Using the recursion relations (\ref{Alpha_flow_1}), (\ref{Mu_flow_1}),  (\ref{D_flow_un}),
and the definition of $g^{U1}$, implies the following recursion relation for $g^{U1}$:
\begin{equation}
	\label{gu1eqn}
	\frac{\dd\ln g^{U1}}{\dd \ell}=\frac{2+2\eta_D+\eta_\alpha-3\eta_\mu}{2} \approx \left({1-g^{U1}}\right),
\end{equation}
where the first equality is exact,
and the second, approximate, equality is valid only to one-loop order, obtained from combining (\ref{gu1def}) with (\ref{Alpha_flow_un_1},\ref{Mu_flow_un_1},\ref{D_flow_un_1}) .
This implies that there is a stable fixed point at $(g^{U1})^* = 1$. While this fixed point is only valid to one-loop order, since $g^{U1}$ \emph{does} have a non-zero  fixed point value at \emph{all} orders, and since the first equality in \eqref{gu1eqn} is valid to \emph{all} orders, setting the L.H.S. of \eqref{gu1eqn} to 0 leads to an \emph{exact} identity between  $\eta_{\alpha,\mu,D}$:
\begin{equation}
2+2\eta_D+\eta_\alpha-3\eta_\mu=0\,.
\end{equation}

An exactly analogous calculation for the second uncontrolled calculation, using the BZ $-\infty<q_x<\infty$, $-\Lambda<q_y<\Lambda$,
obtains the following recursion relations to one-loop order {in exactly $d=2$}:
\begin{eqnarray}
	&&\frac{\dd\ln \alpha}{\dd \ell}
	=\left(z-2 \zeta +2-\frac{7g^{U2}}{8}\right) ,\\
	&&\frac{\dd\ln \mu_x}{\dd \ell}
	=\left(z-2+\frac{5g^{U2}}{8}\right) ,\\
	&&\frac{\dd\ln D }{\dd \ell}
	=\left(z -3+\zeta-\frac{3g^{U2}}{8}\right) ,
\end{eqnarray}
where
\begin{equation}
	g^{U2}=\frac{\zeta\alpha^{1/4}D}{8\sqrt{2}\mu^{5/4}\pi\sqrt{\Lambda}}\,.
\end{equation}
 This definition of $g^{U2}$ implies the recursion relation
\begin{equation}
	\frac{\dd\ln g^{U2}}{\dd \ell}=\frac{2\zeta+4\eta_D+\eta_\alpha-5\eta_\mu}{4} \approx \frac{4\zeta-11 g^{U2}}{8},
\end{equation}
where, again, the first equality is formally valid to all orders, but the second, approximate equality, to only one-loop order.
Since $\zeta>0$, this implies a stable fixed point at $(g^{U2})^* = {4\zeta^*/11}$ to one loop.

Finally, analytically continuing our calculation to $d>2$ by treating the $y$ direction as $d-1$ dimensional, we obtain
\begin{eqnarray}
	&&\frac{\dd\ln \alpha}{\dd \ell}
	=\left(z-2 \zeta +2-\frac{7g^\epsilon}{8}\right)\,,\\
	&&\frac{\dd\ln \mu_x}{\dd \ell}
	=\left(z-2+\frac{5g^\epsilon}{8}\right)\,,\\
	&&\frac{\dd\ln D }{\dd \ell}
	=\left(z -3+(3-d)\zeta-\frac{3g^\epsilon}{8}\right)
\end{eqnarray}
where
\begin{equation}
	g^\epsilon=\frac{\zeta\alpha^{1/4}DS_{d-1}\Lambda^{d-5/2}}{8\sqrt{2}(2\pi)^{d-1}\mu^{5/4}}
\end{equation}
with $\Lambda$ being the large wavenumber cut-off in $q_y$ and the closed recursion relation for $g^\epsilon$ is
\begin{equation}
	\frac{\dd\ln g^\epsilon}{\dd \ell}= \frac{(10-4d)\zeta+4\eta_D+\eta_\alpha-5\eta_\mu}{4}\approx\frac{5-2d}{2}\zeta-\frac{11}{8}g^\epsilon.
\end{equation}
For {$d<{5/2}$}, these flow equations imply a fixed point at $(g^\epsilon)^* = \frac{8\epsilon}{11}\zeta^*$
where $\epsilon=5/2-d$.

We now use the trajectory integral-matching formalism \cite{Nelson_traj} to calculate $C\left(\br, t\right)=\langle\bu(\br,t)\cdot\bu(\mathbf{0},0)\rangle$, by relating the correlators in the original system  to those of the rescaled  system, via
\begin{eqnarray}
	&&C\left(\alpha_0,\mu_{x0},D_{{0}},\br, t\right)\nonumber\\
	&=& e^{2\chi\ell}C\left[\alpha(\ell),\mu_x(\ell),D(\ell),{|y|\over e^{\zeta\ell}},{|x-v_bt|\over e^{\ell}}, {|t|\over e^{z\ell}}\right]\,,~~~~\label{Traje2}
\end{eqnarray}
with $\alpha$, {$\mu_x$} and $D$ controlling the magnitude of $C\left(\br,t\right)$, and the subscript ``0" denotes the bare values of the parameters. Note the argument $x-v_bt$ appears in these expressions because we have undone the aforementioned Galilean boost to return to the laboratory coordinate system.

We will illustrate this calculation in detail for  the first uncontrolled approximation; for  the other approximations the calculation is very similar.
We choose $\zeta$ and $z$ to fix $\alpha$, $\mu_x$, and $D$ (of course, due to the exact relation (\ref{Chi1}), fixing $\zeta$ also fixes $\chi$). This gives
\begin{equation}
z^{U1}={7\over 4}\,, ~~~\zeta^{U1}={3\over 2}\,,~~~\chi^{U1}=-{1\over2}\,.
\end{equation}
Then setting $\ell=\ln\left(\Lambda |x{-v_b t}|\right)$ casts the R.H.S. of \eqref{Traje2} in the form (\ref{Correl1}) where
\begin{eqnarray}
	&&\cG_{_{UET}}\left({|t|\over |x{-v_bt}|^z},{|y|\over |x{-v_bt}|^{\zeta}}\right)\nonumber\\	&\equiv&
\Lambda^{2\chi}C\left[\alpha_0,{\mu_x}_0,D_0,{|y|\over (|x{-v_bt}|\Lambda)^{\zeta}},{1\over\Lambda},{|t|\over (|x{-v_bt}|\Lambda)^z}\right]\,.\nonumber\\ \label{cfn}
\end{eqnarray}
Since the static limit of the correlator $C\left(\br, t=0\right)$ has been obtained in \cite{CLT_Ncomm}, we focus here on the dynamic limit $C\left({\mathbf 0}, t\right)$. In this limit we expect the correlator to be a power law of $t$ only. As a result, the scaling function $\cG_{_{UET}}$ must behave as $\left(t/|x-v_bt|^z\right)^{\chi\over z}$ to cancel out the prefactor $|x-v_bt|^{2\chi}$ in (\ref{Correl1}). Thus we obtain  (\ref{Correl2}). We remind the reader that the static exponents $\zeta^{U1}$ and $\chi^{U1}$ in this approximation, coincidentally, coincide with the exact static exponents obtained in \cite{CLT_Ncomm}. We however note that as these are one-loop calculations, the static exponents have no reason to coincide with their exact values and, indeed, they do not in the other two approximations.

Likewise, the second uncontrolled approximation yields
\begin{equation}
z^{U2}={23\over 14}\,, ~~~\zeta^{U2}={11\over 7}\,,~~~\chi^{U2}=-{4\over 7}\,,
\end{equation}
and the epsilon expansion
\begin{eqnarray}
 z^{\epsilon}&=&2-{10\over 11}\epsilon={17\over 11}\, ,\\
 \zeta^{\epsilon}&=&2- {12\over 11}\epsilon={16\over 11}\,, \\
 \chi^{\epsilon}&=&-1+{12\over 11}\epsilon=-{5\over 11}\,,
\end{eqnarray}
where $\epsilon=5/2-d$ and in the second equality $d=2$ is taken.

Taking an average over the three different sets of values of the exponents
 leads to the value
$z=1.65\pm0.06$
 quoted in the introduction, $\zeta=1.51\pm0.03$, and $\chi=-0.51\pm0.03$. Note that the predictions for $\zeta$ and $\chi$ are extremely
close to their exact values $\zeta=1.5$ and $\chi=-0.50$, which implies our prediction for $z$
is almost certainly also very accurate. This implies that though the
linear theory suggested that velocity fluctuations in incompressible flocks are diffusive,
nonlinearities \emph{lower} the dynamical exponent, which  makes the
dynamics \emph{super}diffusive: that is, distance scales like $t^{1/z}$ with $z<2$, which
means distance grows faster with time than the $t^{1/2}$ diffusive law.

In summary, we have calculated the dynamical exponent $z$ characterizing the scaling of velocity fluctuations with time in
incompressible two-dimensional flocks.  We did so using \emph{three} different approximation schemes,  and demonstrated that the dynamics is significantly modified; specifically,  the value of $z$ is substantially changed  by nonlinearities, which turn the fluctuations superdiffusive. Interestingly, due to the mapping \cite{CLT_Ncomm} between this system and the model-A dynamics of a divergence-free magnet \cite{Kashuba}, this also implies that magnetisation fluctuations of a divergence-free  magnet or a two-dimensional magnet with two-dimensional dipolar interactions are also superdiffusive.

\begin{acknowledgments}
L.C. acknowledges support by the National Science
Foundation of China (under Grant No. 11874420), and thanks the MPI-PKS, where the early stage of this work was performed, for their support. J. T.
 thanks the Max Planck Institute for the Physics of Complex Systems,
Dresden, Germany, for their support through a Martin
Gutzwiller Fellowship during this period. AM was supported by a TALENT fellowship from CY Cergy Paris Universit\'e.
\end{acknowledgments}

\clearpage
\title{Dynamics of packed swarms: time-displaced correlators of two dimensional incompressible flocks\\ Supplementary material}
\address{School of Material Science and Physics, China University of Mining and Technology, Xuzhou Jiangsu, 221116, P. R. China}
\address{Department of Bioengineering, Imperial College London, South Kensington Campus, London SW7 2AZ, U.K.}
\address{Laboratoire de Physique Th\'eorique et Mod\'elisation, CNRS UMR 8089,
	CY Cergy Paris Universit\'e, F-95032 Cergy-Pontoise Cedex, France}
\address{{Sorbonne Universit\'{e} and CNRS, Laboratoire Jean Perrin, F-75005, Paris, France}}
\affiliation{Department of Physics and Institute of Theoretical
	Science, University of Oregon, Eugene, OR $97403^1$}
\affiliation{Max Planck Institute for the Physics of Complex Systems, N\"othnitzer Str. 38, 01187 Dresden, Germany}
\maketitle
\onecolumngrid
\appendix
\setcounter{equation}{0}
\section{Review of the DRG procedure}
Our approach is that developed by \cite{forster_pra77}, to which the reader is referred for a more detailed discussion.

We start with the EOM in Fourier space:
\beq
-\ii\omega u_y(\tilde{\bq})=P_{yx}\left(\bq\right)\mathcal{F}_{\tilde{\bq}}
\left[-\alpha\left(u_x+{u_y^2\over 2}\right)\right]+P_{yy}\left(\bq\right)\mathcal{F}_{\tilde{\bq}}
\left[-\alpha
\left(u_x+{u_y^2\over 2}\right)u_y+\mu_x\partial_x^2u_y+f_y\right]\,,
\label{SM_2Dy4}
\eeq
where
\beq
\tilde{\bq}\equiv (\omega, \bq)\,,~~~P_{yy}(\bq)=q_x^2/q^2\,,~~~ P_{xy}(\bq)=-q_xq_y/q^2\,.
\eeq
We can solve (\ref{SM_2Dy4}) for $u_y(\tilde{\bq})$ to get
\beqn
u_y(\tilde{\bq})&=&
G(\tilde{\bq}) \Bigg[
P_{yy}(\bq) f_y(\tilde{\bq})
+\left(\frac{\alpha}{2}\right) P_{yx} (\bq) \int_{\tilde{\bk}}
u_y(\tilde{\bk})u_y(\tilde{\bq}-\tilde{\bk})
-\alpha P_{yy} (\bq) \int_{\tilde{\bk}}
u_y(\tilde{\bk})u_x(\tilde{\bq}-\tilde{\bk})\nonumber
\\
&&
-\left(\frac{\alpha}{2}\right) P_{yy} (\bq) \int_{\tilde{\bk},\tilde{\bk}'}
u_y(\tilde{\bk})u_y(\tilde{\bk}')u_y(\tilde{\bq}-\tilde{\bk}-\tilde{\bk}')
\Bigg]\,,\label{SM_EOM_1}
\eeqn
where
\beqn
&&\tilde{\bk}\equiv (\Omega, \bk)\,,~~~\tilde{\bk}'\equiv (\Omega', \bk')\,,\\
&&\int_{\tilde{\bk}}\equiv \int_{-\infty}^{\infty}{\dd\Omega\over\sqrt{2\pi}}\int{\dd^dk\over (2\pi)^{d/2}}\,,
~~~
\int_{\tilde{\bk}'}\equiv \int_{-\infty}^{\infty}{\dd\Omega'\over \sqrt{2\pi}}\int{\dd^dk'\over (2\pi)^{d/2}}\,,\\
\eeqn
where the integrals of $\bk$ and $\bk'$ are over the entire Brillouin zone. The propagator $G(\tilde{\bq})$ is defined as:
\beq
G(\tilde{\bq})=
\frac{1}{-\ii \omega +\mu_x q_x^2+\mu_\perp q_y^2 +\alpha \frac{q_y^2}{q^2}}
\approx
\frac{1}{-\ii \omega +\mu_x q_x^2 +\alpha \frac{q_y^2}{q_x^2}}
\ ,
\eeq
where the second, approximate, equality   is asymptotically exact in the long wave length limit $q\to 0$.

The statistics of the random force $f_y$ in Fourier space are given by
\beq
\la f_y(\tilde{\bq}) f_y(\tilde{\bq}')\ra=2D \delta^d(\bq+\bq') \delta(\omega+\omega') \ .
\eeq

In the linear approximation, Eq. (\ref{SM_EOM_1}) reduces to
\beq
u_y(\tilde{\bq})=G(\tilde{\bq}) P_{yy}(\bq) f_y(\tilde{\bq})\,.
\eeq
Autocorrelating both sides of (\ref{SM_EOM_Lin_1})  we get
\beq
\langle u_y(\tilde{\bq})u_y(\tilde{\bq}')\rangle=2D\left(q_{x}^2\over q^2\right){C(\tilde{\bq})\delta(\omega+\omega')\delta(\bq+\bq') }\,,\label{SM_Lin_uy}
\eeq
where
\beq
C(\tilde{\bq}) = \frac{  2D}{\omega^2+\left(\mu_x q_x^2+ \alpha  {q_y^2\over q_x^2}\right)^2}
\ .\label{SM_EOM_Lin_1}
\eeq
Combining this with the incompressibility constraint $u_x(\tilde{\bq})=-q_yu_y(\tilde{\bq})/q_x$
leads to the  correlators for other components of $\bu(\tilde{\bq})$:
\begin{subequations}
	\begin{align}
		\langle u_x(\tilde{\bq})u_x(\tilde{\bq}')\rangle=2D\left(q_{y}^2\over q^2\right){C(\tilde{\bq})\delta(\omega+\omega')\delta(\bq+\bq') },\label{SM_Lin_ux}\\
		\langle u_x(\tilde{\bq})u_y(\tilde{\bq}')\rangle=-2D\left(q_xq_y\over q^2\right)C(\tilde{\bq}){\delta(\omega+\omega')\delta(\bq+\bq')}\,.
	\end{align}
\end{subequations}
Adding (\ref{SM_Lin_uy}) and (\ref{SM_Lin_ux}) together, we get the full auto-correlator of $\bu$:
\beq
\langle \bu(\tilde{\bq})\cdot\bu(\tilde{\bq}')\rangle=2D{C(\tilde{\bq})\delta(\omega+\omega')\delta(\bq+\bq') }\,.\label{}
\eeq

We see that in the low frequency and small wavenumber limit (i.e., $\omega\to 0$, $q\to 0$) $\langle \bu(\tilde{\bq})\cdot\bu(\tilde{\bq}')\rangle$ diverges most strongly in the regime $\omega\sim q_y\sim q_x^2$. In this dominant regime, $P_{yy}(\bq)$ can be approximated as 1. Therefore, the EOM (\ref{SM_EOM_1}) can be simplified as
\beqn
u_y(\tilde{\bq})&=&
G(\tilde{\bq}) \Bigg[
f_y(\tilde{\bq})
+\left(\frac{\alpha}{2}\right) P_{yx} (\bq) \int_{\tilde{\bk}}
u_y(\tilde{\bk})u_y(\tilde{\bq}-\tilde{\bk})
-\alpha \int_{\tilde{\bk}}
u_y(\tilde{\bk})u_x(\tilde{\bq}-\tilde{\bk})\nonumber
\\
&&
-\left(\frac{\alpha}{2}\right) \int_{\tilde{\bk},\tilde{\bk}'}
u_y(\tilde{\bk})u_y(\tilde{\bk}')u_y(\tilde{\bq}-\tilde{\bk}-\tilde{\bk}')
\Bigg]\,.\label{SM_EOM_2}
\eeqn

\begin{figure}
	\begin{center}
		\includegraphics[scale=.4]{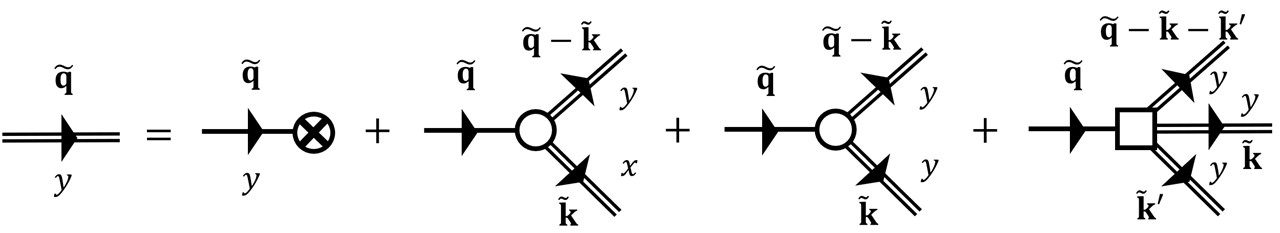}
	\end{center}
	\caption{The graphical representation of the EOM (\ref{SM_EOM_2}).}
	\label{fig:EOM}
\end{figure}

\begin{figure}
	\begin{center}
		\includegraphics[scale=.4]{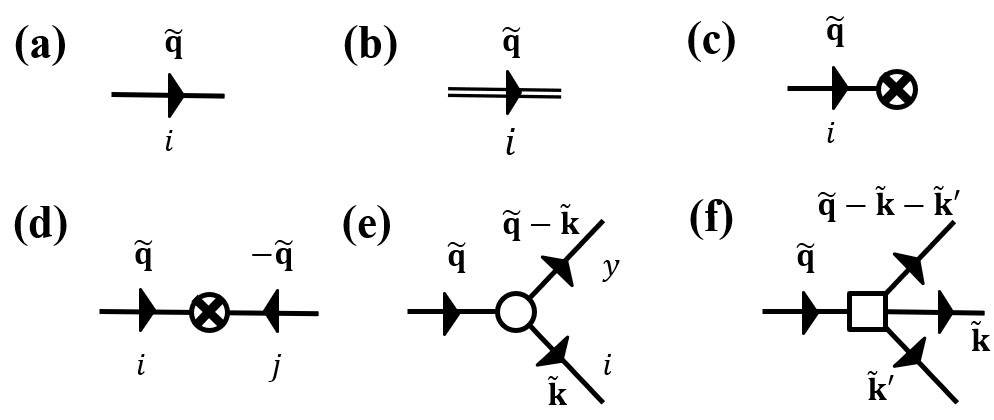}
	\end{center}
	\caption{Definitions of the various elements in the Feynman diagrams:
		(a) $={q_y\left(1-2\delta_{ix}\right)\over q_i}G\left(\tilde{\bq}\right) $, (b) $=u_i(\tilde{\bq})$, (c) $={q_y\left(1-2\delta_{ix}\right)\over q_i}G\left(\tilde{\bq}\right) f_y(\tilde{\bq})$,
		(d) $=P_{ij}(\bq)C(\tilde{\bq})$,
		(e) ${\rm{the~circle}}=-\alpha\left(1-{1\over 2}\delta_{yi}\right)\left[\delta_{xi}
		+P_{yx}\left(\bq\right)\delta_{yi}\right]$,
		(f) the square $= -\alpha/2$.}
	\label{fig:Element}
\end{figure}

\begin{figure}
	\begin{center}
		\includegraphics[scale=.2]{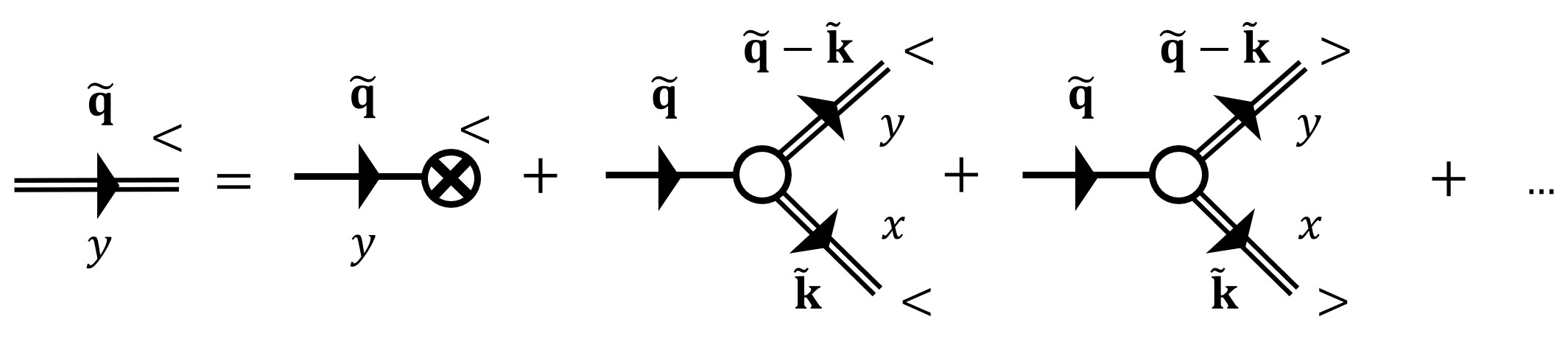}
	\end{center}
	\caption{The graphical representation of the solution for $u^<(\tilde{\bq})$.}
	\label{fig:EOM_Slow}
\end{figure}

\begin{figure}
	\begin{center}
		\includegraphics[scale=.2]{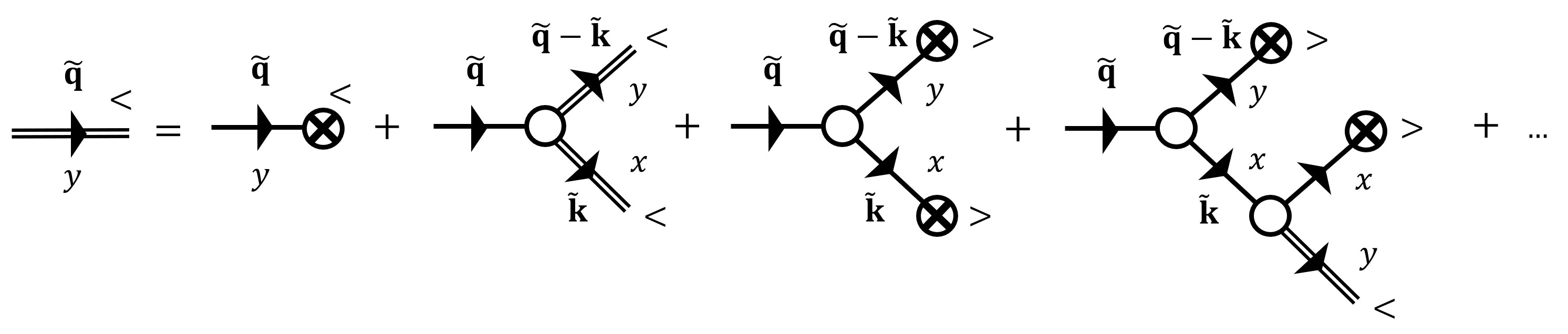}
	\end{center}
	\caption{The graphical representation of the solution for $u^<(\tilde{\bq})$ after the iteration of $u^>(\tilde{\bq})$.}
	\label{fig:EOM_Iterate}
\end{figure}

\begin{figure}
	\begin{center}
		\includegraphics[scale=.5]{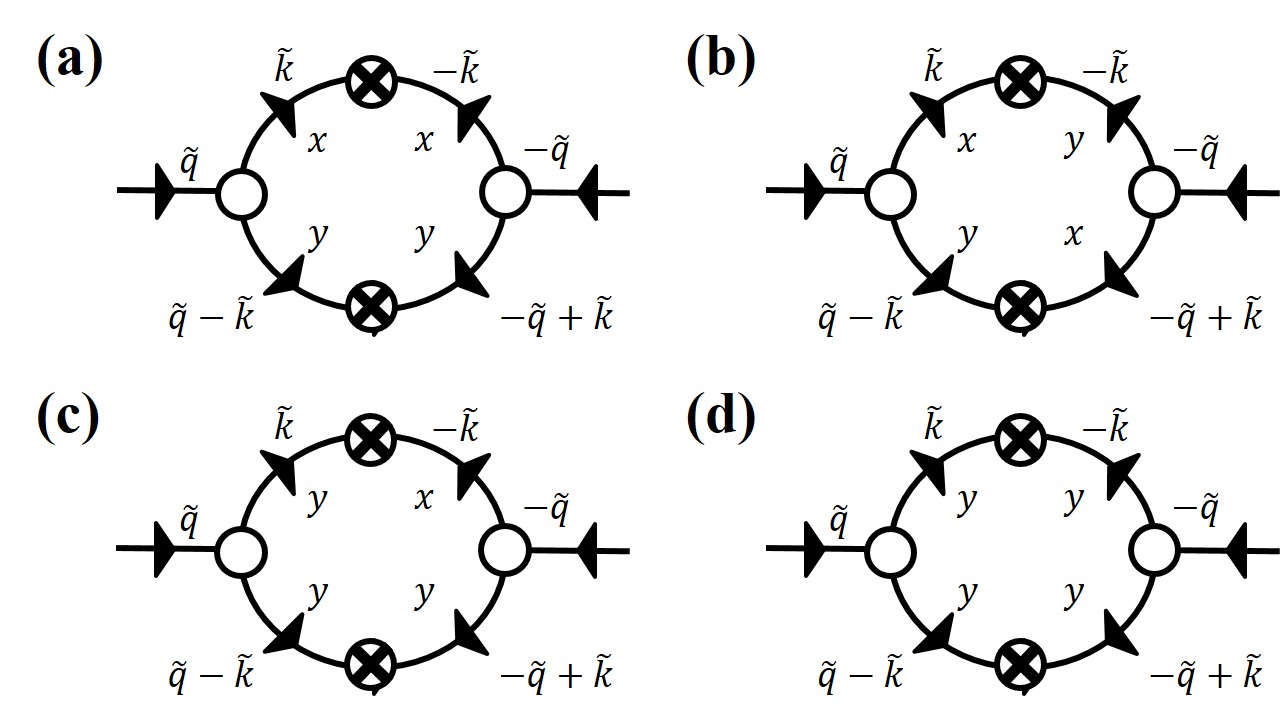}
	\end{center}
	\caption{The graphical representations of the correction to $D$.}
	\label{fig:noise}
\end{figure}

\begin{figure}
	\begin{center}
		\includegraphics[scale=.5]{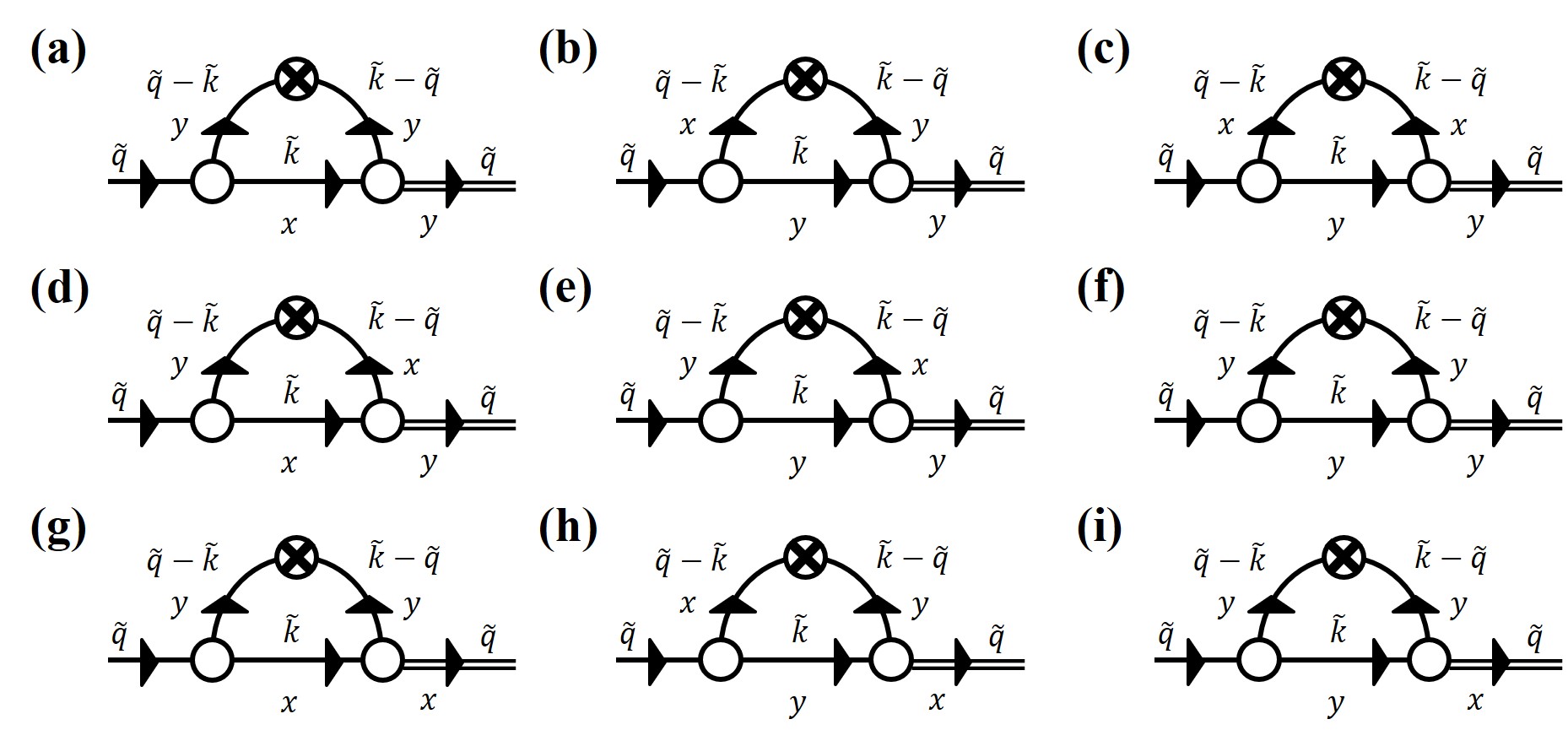}
	\end{center}
	\caption{The graphical representations of the correction to the propagator.}
	\label{fig:propagator}
\end{figure}

Now we perform the perturbative DRG. First we partition the field $\bu(\tilde{\bq})$ into ``slow" and ``fast" parts $\bu^<(\tilde{\bq})$ and $\bu^>(\tilde{\bq})$, which are supported on small $\bq$'s  and large $\bq$'s, respectively. This can be represented by Feynman diagrams. The solutions for $u_y(\bq)$  (\ref{SM_EOM_2})  and for $u_y^<(\bq)$ after the decomposition are represented by the graphs in Figs. \ref{fig:EOM} and \ref{fig:EOM_Slow}, respectively, and the meanings of the various elements of the graphs are depicted in Fig. \ref{fig:Element}.
Next we iteratively solve for  $\bu^>(\tilde{\bq})$. Then we average out all the short wave length (i.e., ``$>$'') fluctuations. For instance, the third graph on the RHS in Fig. \ref{fig:EOM_Iterate} represents a contribution to the large wave length noise $G(\tilde{\bq})f_y^<(\tilde{\bq})$. The correction to the noise strength can be calculated by autocorrelating this contribution, which is represented by the graphs (a) and (b) in Fig. \ref{fig:noise}. Also, for the fourth graph on the RHS in Fig. \ref{fig:EOM_Iterate} we do a pairwise average over the short wave length noise. This is represented by connecting the two legs labeled with ``$>$''. The resultant graph is just the one (d) in Fig. \ref{fig:propagator}, representing a contribution to the terms linear in $\bu^<(\tilde{\bq})$. Following this procedure we get all the other one-loop graphs listed in Figs. \ref{fig:noise} and \ref{fig:propagator}.

Eventually, eliminating $\bu^>(\tilde{\bq})$ leads to an effective EOM for $\bu^<(\tilde{\bq})$,
\beqn
u^<_y(\tilde{\bq})&=&
G(\tilde{\bq}) \Bigg[
f^<_y(\tilde{\bq})+\delta  f^<_y(\tilde{\bq})+\ii\omega\left(\eta_\omega\dd\ell\right)  u^<_y(\tilde{\bq})-\alpha\left(\eta_\alpha^{dir}\dd\ell\right) P_{yx}(\bq)u_x^<(\tilde{\bq})-\mu_x\left(\eta_\mu^{dir}\dd\ell\right) q_x^2 u^<_y(\tilde{\bq})\Bigg]\nonumber\\
&&+G(\tilde{\bq}) \Bigg\{\alpha\left(1+\eta_\alpha^{dir}\dd\ell\right) \mbox{NL}\left[\bu^<(\tilde{\bq})\right]\Bigg\}\,,\label{SM_EOM_3}
\eeqn
where $\dd\ell$ is an infinitesimal number quantifying the thickness of the momentum shell occupied by the large wavevectors, and $\mbox{NL}\left[\bu^<(\tilde{\bq})\right]$ stands for the nonlinear part:
\beqn
\mbox{NL}\left[\bu(\tilde{\bq})\right]=
\left(\frac{1}{2}\right) P_{yx} (\bq) \int_{\tilde{\bk}}
u_y(\tilde{\bk})u_y(\tilde{\bq}-\tilde{\bk})
- \int_{\tilde{\bk}}
u_y(\tilde{\bk})u_x(\tilde{\bq}-\tilde{\bk})
-\left(\frac{1}{2}\right) \int_{\tilde{\bk},\tilde{\bk}'}
u_y(\tilde{\bk})u_y(\tilde{\bk}')u_y(\tilde{\bq}-\tilde{\bk}-\tilde{\bk}')\,.\label{}
\eeqn
The autocorrelation of the noise contribution $\delta f^<_y(\tilde{\bq})$ is given by
\beq
\langle 2\delta f^<_y(\tilde{\bq})\delta f^<_y(\tilde{\bq}')\rangle=2\left(1+\eta_D^{dir}\dd\ell\right)D\delta^d(\bq+\bq')\delta(\omega+\omega')\,.
\eeq
The detailed calculation of $\eta_{\alpha,D,\mu}^{dir}$ to one-loop order is given in Sec. \ref{Sec:graphical correction}. 

We multiply both sides of (\ref{SM_EOM_3}) by $\left[G(\tilde{\bq})\right]^{-1}$
and rearrange the terms to get
\beqn
-\ii\omega(1+\eta_\omega\dd\ell)u^<_y(\tilde{\bq})&=&
\Bigg[
f'_y(\tilde{\bq}) -\alpha\left(1+\eta_\alpha^{dir}\dd\ell\right) P_{yx}(\bq)u_x^<(\tilde{\bq})-\mu_x\left(1+\eta_\mu^{dir}\dd\ell\right) q_x^2 u^<_y(\tilde{\bq})\Bigg]\nonumber\\
&&+\Bigg\{\alpha\left(1+\eta_\alpha^{dir}\dd\ell\right) \mbox{NL}\left[\bu^<(\tilde{\bq})\right]\Bigg\}\,.\label{SM_EOM_4}
\eeqn
Then we divide both sides of (\ref{SM_EOM_4}) by $(1+\eta_\omega\dd\ell)$. This gives
\beqn
-\ii\omega u^<_y(\tilde{\bq})&=&
\Bigg[
{f'_y(\tilde{\bq})\over 1+\eta_\omega\dd\ell} -{\alpha\left(1+\eta_\alpha^{dir}\dd\ell\right) \over 1+\eta_\omega\dd\ell} P_{yx}(\bq)u_x^<(\tilde{\bq})-{\mu_x\left(1+\eta_\mu^{dir}\dd\ell\right) \over 1+\eta_\omega\dd\ell}q_x^2 u^<_y(\tilde{\bq})\Bigg]\nonumber\\
&&+\Bigg\{{\alpha\left(1+\eta_\alpha^{dir}\dd\ell\right) \over 1+\eta_\omega\dd\ell} \mbox{NL}\left[\bu^<(\tilde{\bq})\right]\Bigg\}\,,\label{SM_EOM_5}
\eeqn
which is essentially (\ref{SM_EOM_2}) with ``renormalized" noise strength and coefficients:
\beqn
&&D_r={\left(1+\eta_D^{dir}\dd\ell\right)D\over (1+\eta_\omega\dd\ell)^2}=\left[1+\left(\eta_D^{dir}-2\eta_\omega\right)\dd\ell\right]D\,,\label{SM_Re_D_1}\\
&&\alpha_r={\left(1+\eta_\alpha^{dir}\dd\ell\right)\alpha\over 1+\eta_\omega\dd\ell}=\left[1+\left(\eta_\alpha^{dir}-\eta_\omega\right)\dd\ell\right]\alpha\,,\\
&&\mu_x^r={\left(1+\eta_\mu^{dir}\dd\ell\right)\mu_x^r\over 1+\eta_\omega\dd\ell}=\left[1+\left(\eta_\mu^{dir}-\eta_\omega\right)\dd\ell\right]\mu_x\,.
\eeqn
where
\begin{equation}
	\eta_D=\eta^{dir}_D-2\eta_\omega\,;~~~\eta_\alpha=\eta^{dir}_\alpha-\eta_\omega;\, ~~~\eta_\mu=\eta^{dir}_\mu-\eta_\omega\,.
	. \label{SM_Gr_Cor_1}
\end{equation}

Note that the graphical correction for each parameter has two parts: the $\eta$ with superscript ``dir" and $\eta_\omega$ (or 2$\eta_\omega$ in the case of $D$). The former comes from the ``direct" renormalization of the parameter in the process of eliminating $\bu^>(\tilde{\bq})$, while the latter comes from rescaling the renormalized coefficient of $-\ii\omega u_y^<(\tilde{\bq})$ back to 1. Furthermore, the graphical corrections to the three $\alpha$'s are identical. This is not a coincidence, but rather due to the symmetry (i.e., rotation invariance) of (\ref{SM_2Dy4}), which must be preserved
under the DRG transformation.
	This symmetry argument implies that we do not have to explicitly calculate graphical corrections to the quadratic and cubic vertices themselves, thus simplifying our calculation considerably.
	
	Also note that EOM (\ref{SM_2Dy4})  can be written as a detailed-balance-obeying dynamics of the form
	\begin{equation}
		\partial_t u_l=-\Gamma\frac{\delta \mathcal{H}}{\delta u_l}+f_l
	\end{equation}
	where $\mathcal{H}$ is the Hamiltonian for a divergence-free magnet, and the friction coefficient $\Gamma$ is taken to be unity. This dynamics obeys fluctuation-dissipation theorem which \emph{must} be preserved under renormalization. This imposes  a  constraint: the ratio of the remormalized noise strength to the renormalized friction coefficient is fixed 
The  renormalized friction coefficient appears as the common factor $1/(1+\eta_\omega\dd\ell)$ on the RHS in (\ref{SM_EOM_5}). The renormalized noise strength is given by (\ref{SM_Re_D_1}). Then it follows from the constraint imposed by fluctuation-dissipation theorem that
\beq
\eta_\omega=\eta_D^{dir}\,. \label{SM_Flu_Dis_1}
\eeq
Inserting (\ref{SM_Flu_Dis_1}) into (\ref{SM_Gr_Cor_1}) we get
\beqn
\eta_D=-\eta^{dir}_D\,,\label{SM_Gr_Cor_2}\\
\eta_\alpha=\eta^{dir}_\alpha-\eta_D^{dir}\,,\label{SM_Alpha_Cor_1}\\
\eta_\mu=\eta^{dir}_\mu-\eta_D^{dir}\,.\label{SM_Mu_Cor_1}
\eeqn
That is, because of  (\ref{SM_Flu_Dis_1}) , we do not need to explicitly calculate the graphical corrections to $-i\omega u^<_y(\tilde{\bq})$.

Finally we rescale time, lengths, and  fields as
\beqn
t\to te^{z\dd\ell},~~x\to xe^{\dd\ell},~~y\to ye^{\zeta\dd\ell},\\
u_y\to u_ye^{\chi\dd\ell},~~u_x\to u_xe^{\left(\chi+1-\zeta\right)\dd\ell}\,,
\eeqn
to restore  the Brillouin zone to its original size.


To take further the advantage of rotational invariance of the EOM, we choose  $\chi = 1-\zeta$ so that the three $\alpha$'s remain identical upon renormalization. This relation is, unsurprisingly, obeyed by their exact values $\chi =-1/2$ and $\zeta=3/2$.

So far we have completed one cycle of DRG procedure. Repeating this cycle ad infinitum leads to the recursion relations for $D$, $\alpha$, and $\mu_x$ given in the main text.


	%


%

\section{Calculation of graphical corrections\label{Sec:graphical correction}}
\subsection{Direct noise correction}
According to the rules illustrated in Fig. \ref{fig:Element}, the graphs in Fig.~\ref{fig:noise} give contributions to the autocorrelation of $G(\tilde{\bq})f^<_y(\tilde{\bq})$. Specifically, graph (a) gives
\begin{equation}
	\alpha^2  G(\tilde{\bq})G(-\tilde{\bq})
	\int{\dd\Omega\over 2\pi}\int{\dd^dk\over (2\pi)^d}
	\frac{k_y^2}{k^2} \frac{(q_x-k_x)^2}{|\bq-\bk|^2} C(\tilde{\bk}) C(\tilde{\bq}-\tilde{\bk}) \,,
	\label{SM_Noise_1}
\end{equation}
where $\int_>\dd^dk$ integrates over large $\bk$ (i.e., the $\bk$ in the momentum shell); graph (b) gives
\begin{equation}
	\alpha^2  G(\tilde{\bq})G(-\tilde{\bq})
	\int{\dd\Omega\over 2\pi}\int{\dd^dk\over (2\pi)^d}
	\frac{k_xk_y}{k^2} \frac{(q_x-k_x)(q_y-k_y)}{|\bq-\bk|^2}C(\tilde{\bk}) C(\tilde{\bq}-\tilde{\bk}) \,;
	\label{SM_Noise_2}
\end{equation}
graph (c) gives
\begin{equation}
	-2\alpha^2  P_{xy}(\bq)G(\tilde{\bq})G(-\tilde{\bq})
	\int{\dd\Omega\over 2\pi}\int{\dd^dk\over (2\pi)^d}
	\frac{k_xk_y}{k^2} \frac{(q_x-k_x)^2}{|\bq-\bk|^2}C(\tilde{\bk}) C(\tilde{\bq}-\tilde{\bk}) \,;
	\label{SM_Noise_3}
\end{equation}
graph (d) gives
\begin{equation}
	2\alpha^2  \left[P_{xy}(\bq)\right]^2G(\tilde{\bq})G(-\tilde{\bq})
	\int{\dd\Omega\over 2\pi}\int{\dd^dk\over (2\pi)^d}
	\frac{k_x^2}{k^2} \frac{(q_x-k_x)^2}{|\bq-\bk|^2}C(\tilde{\bk}) C(\tilde{\bq}-\tilde{\bk}) \,;
	\label{SM_Noise_4}
\end{equation}

Since we are interested in the contributions to the white noise, we can set the wavevector $\bq$ inside the loop integral to zero. Since the prefactors $P_{xy}(\bq)$ and $\left[P_{xy}(\bq)\right]^2$ are both much less than 1 in the dominant region $q_y\sim q_x^2$, the contributions (\ref{SM_Noise_3}) and (\ref{SM_Noise_4}) are negligible compared to (\ref{SM_Noise_1}) and (\ref{SM_Noise_2}).

Upon setting set the wavevector $\bq$ inside the loop integral to zero, (\ref{SM_Noise_1}) and (\ref{SM_Noise_2}) become identical, and their sum is
\begin{equation}
	2\alpha^2  G(\tilde{\bq})G(-\tilde{\bq})
	\int {\dd\Omega\over 2\pi}\int{\dd^dk\over (2\pi)^d}
	\frac{k_y^2}{k_x^2} C(\tilde{\bk}) C(-\tilde{\bk}) =
	2\alpha^{2} D^2 G(\tilde{\bq})G(-\tilde{\bq})
	\int{\dd^dk\over (2\pi)^d}
	\frac{k_y^2k_x^4}{\left( \mu k_x^4
		+\alpha k_y^2\right)^3}\,,
\end{equation}
which implies the following correction to the noise strength:
\begin{equation}
	\delta D^{dir}=\alpha^{2} D^2  I_1\,,
\end{equation}
where
\begin{equation}
	I_1=\int{\dd^dk\over (2\pi)^d}
	\frac{k_y^2k_x^4}{\left( \mu k_x^4
		+\alpha k_y^2\right)^3}\,.
\end{equation}

Using the results for $I_1$ calculated in Sec. \ref{Sec: Integrals} we get
\beqn
\mbox{Uncontrolled approximations with}~ k_x\in \{-\infty, \infty\}:~~~\delta D^{dir}=\frac{3}{8} D g^u_\parallel \dd\ell\,,\\
\mbox{Uncontrolled approximations with}~ k_y\in \{-\infty, \infty\}:~~~\delta D^{dir}=\frac{1}{4} D g^u_\perp \dd\ell\,,\\
\mbox{$\epsilon$ expansion with}~ k_x\in \{-\infty, \infty\}:~~~\delta D^{dir}=\frac{3}{8} D g_h \dd\ell\,,
\eeqn
where
\begin{equation}
	g^u_\parallel=\frac{\zeta\alpha^{1/4}D}{8\sqrt{2}\mu^{5/4}\pi\sqrt{\Lambda}}\,,~~~
	g^u_\perp=\frac{\alpha^{1/2}D}{4\mu^{3/2}\pi{\Lambda}}\,,~~~
	g_h=\frac{\zeta\alpha^{1/4}DS_{d-1}\Lambda^{d-5/2}}{8\sqrt{2}(2\pi)^{d-1}\mu^{5/4}}\,.
\end{equation}
This implies
\beqn
\mbox{Uncontrolled approximations with}~ k_x\in \{-\infty, \infty\}:~~~\eta_D^{dir}=\frac{3}{8}  g^u_\parallel\,,\label{SM_D_Dir_Ux}\\
\mbox{Uncontrolled approximations with}~ k_y\in \{-\infty, \infty\}:~~~\eta_D^{dir}=\frac{1}{4}  g^u_\perp\,,\label{SM_D_Dir_Uy}\\
\mbox{$\epsilon$ expansion with}~ k_x\in \{-\infty, \infty\}:~~~\eta_D^{dir}=\frac{3}{8} g_h\,.\label{SM_D_Dir_Epsilon}
\eeqn

Inserting (\ref{SM_D_Dir_Ux}), (\ref{SM_D_Dir_Uy}), and (\ref{SM_D_Dir_Epsilon}) into (\ref{SM_Gr_Cor_2}), respectively, we obtain
\beqn
\mbox{Uncontrolled approximations with}~ k_x\in \{-\infty, \infty\}:~~~\eta_D=-\frac{3}{8}  g^u_\parallel\,,\\
\mbox{Uncontrolled approximations with}~ k_y\in \{-\infty, \infty\}:~~~\eta_D=-\frac{1}{4}  g^u_\perp\,,\\
\mbox{$\epsilon$ expansion with}~ k_x\in \{-\infty, \infty\}:~~~\eta_D=-\frac{3}{8} g_h\,.
\eeqn

\subsection{Direct $\alpha$ correction}
The only graphical contribution to the term $G(\tilde{\bq})P_{yx}(\bq)u_x^<(\tilde{\bq})$ in (\ref{SM_EOM_3}) comes from graph  (i)  in Fig. \ref{fig:propagator}, which corresponds to
\beq
\alpha^2  G(\tilde{\bq}) P_{xy}(\bq)u_x^<(\tilde{\bq})\int {\dd\Omega\over 2\pi}\int{\dd^dk\over (2\pi)^d} \frac{k_x^2}{k^2}
G(\tilde{\bk}) C(\tilde{\bq}-\tilde{\bk})\,.
\eeq

Setting both $\omega$ and $\bq$ to zero inside the integrals and integrating over $\Omega$, we have 
\begin{equation}
	\frac{\alpha^2  D}{2}G(\tilde{\bq}) P_{xy}(\bq)u_x^<(\tilde{\bq})
	\int{\dd^dk\over (2\pi)^d}
	\frac{{k_x^6}}{(\mu k_x^4+\alpha k_y^2)^2k^2}\,.\label{SM_Alpha_1}
\end{equation}
Noting that the integral over the region $k_x^2/k^2\approx 1$ dominates, we can rewrite (\ref{SM_Alpha_1}) as
\begin{equation}
	\frac{\alpha^2  D}{2}G(\tilde{\bq}) P_{xy}(\bq)u_x^<(\tilde{\bq})
	\int{\dd^dk\over (2\pi)^d}
	\frac{{k_x^4}}{(\mu k_x^4+\alpha k_y^2)^2}\,,
\end{equation}
which implies
\begin{equation}
	\delta\alpha^{dir}=-\frac{\alpha^2  D}{2}I_2\,,
\end{equation}
where $I_2$ is defined as
\beq
I_2=\int{\dd^dk\over (2\pi)^d}
\frac{{k_x^4}}{(\mu k_x^4+\alpha k_y^2)^2}\,.
\eeq

Using the results for $I_2$ calculated in Sec. \ref{Sec: Integrals} we get
\beqn
\mbox{Uncontrolled approximations with}~ k_x\in \{-\infty, \infty\}:~~~\delta \alpha^{dir}=-{1\over 2}\alpha g^u_\parallel \dd\ell\,,\\
\mbox{Uncontrolled approximations with}~ k_y\in \{-\infty, \infty\}:~~~\delta \alpha^{dir}= -{1\over 2}\alpha g^u_\perp \dd\ell\,,\\
\mbox{$\epsilon$ expansion with}~ k_x\in \{-\infty, \infty\}:~~~\delta \alpha^{dir}=-{1\over 2} \alpha g_h \dd\ell\,,
\eeqn
which implies
\beqn
\mbox{Uncontrolled approximations with}~ k_x\in \{-\infty, \infty\}:~~~\eta_\alpha^{dir}=-{1\over 2}g^u_\parallel\,,\label{SM_Alpha_Dir_Ux}\\
\mbox{Uncontrolled approximations with}~ k_y\in \{-\infty, \infty\}:~~~\eta_\alpha^{dir}=-{1\over 2}g^u_\perp\,,\label{SM_Alpha_Dir_Uy}\\
\mbox{$\epsilon$ expansion with}~ k_x\in \{-\infty, \infty\}:~~~\eta_\alpha^{dir}= -{1\over 2}g_h\,.\label{SM_Alpha_Dir_Epsilon}
\eeqn

Inserting (\ref{SM_Alpha_Dir_Ux}), (\ref{SM_Alpha_Dir_Uy}), and (\ref{SM_Alpha_Dir_Epsilon}) into (\ref{SM_Alpha_Cor_1}), respectively, we obtain
\beqn
\mbox{Uncontrolled approximations with}~ k_x\in \{-\infty, \infty\}:~~~\eta_\alpha=-\frac{7}{8}  g^u_\parallel\,,\\
\mbox{Uncontrolled approximations with}~ k_y\in \{-\infty, \infty\}:~~~\eta_\alpha=-\frac{3}{4}  g^u_\perp\,,\\
\mbox{$\epsilon$ expansion with}~ k_x\in \{-\infty, \infty\}:~~~\eta_\alpha=-\frac{7}{8} g_h\,.
\eeqn

\subsection{Direct $\mu$ correction}
The graphical contribution to the term $G(\tilde{\bq})q_x^2u_y^<(\tilde{\bq})$ in (\ref{SM_EOM_3}) comes from graphs  (a), (b), (c), and (d) in Fig. \ref{fig:propagator}. Note that in the previous two calculations how $\bq$ splits  between the two legs inside the loop does {\it not} affect the results since we set $\bq={\mathbf 0}$. It {\it does} now since the loop integrals have to be expanded to $O(q_x^2)$ to get the factor $q_x^2$. We make the choice that $\bq$ splits equally on the two internal legs. The same choice was made in the seminal paper for ``randomly stirred incompressible fluids"\cite{forster_pra77}.
For notational convenience we define the wavevectors on the two internal leg as the follows:
\beq
{\bf p}={{\bf q}\over 2}+{\bf k}\,, ~~~{\bf h}={{\bf q}\over 2}-{\bf k}\,.
\eeq

Then graph (a) in Fig. \ref{fig:propagator} can be expressed as
\begin{equation}
	\alpha^2  G(\tilde{\bq}) u_y(\tilde{\bq})\int {\dd\Omega\over 2\pi}\int{\dd^dk\over (2\pi)^d}
	\left(\frac{p_y}{p_x}\right)^2G(\tilde{\bp}) C(\tilde{\bh})\,.
\end{equation}
We set  $\omega$ and $q_y$ to zero inside the integral and integrate over $\Omega$. Next we expand the resultant formula to quadratic order in $q_x$ and focus on the $q_x^2$ piece. This gives  the following contribution to $G(\tilde{\bq})q_x^2u_y^<(\tilde{\bq})$ in (\ref{SM_EOM_3}):
\begin{equation}
	\frac{\alpha^2  D}{8} G(\tilde{\bq}) q_x^2u_y(\tilde{\bq})\int{\dd^dk\over (2\pi)^d}
	\frac{\left(5 \alpha^2 k_y^6-10 \alpha \mu k_x^4 k_y^4 +\mu^2 k_x^8k_y^2 \right)}{\left(\alpha k_y^2+\mu k_x^4 \right)^4}=	\frac{\alpha^2  D}{8} G(\tilde{\bq}) q_x^2u_y(\tilde{\bq})(5I_3-10I_4+I_5)\,,\label{SM_Graph_a}
\end{equation}
where
\beqn
I_3=\int{\dd^dk\over (2\pi)^d}	\frac{ \alpha^2 k_y^6 }{\left(\alpha k_y^2+\mu k_x^4 \right)^4}\,,\\
I_4=\int{\dd^dk\over (2\pi)^d}	\frac{\alpha \mu k_x^4 k_y^4 }{\left(\alpha k_y^2+\mu k_x^4 \right)^4},\\
I_5=\int{\dd^dk\over (2\pi)^d}	\frac{\mu^2 k_x^8k_y^2}{\left(\alpha k_y^2+\mu k_x^4 \right)^4}\,.
\eeqn

Similarly, graph (b) in Fig. \ref{fig:propagator} represents
\begin{equation}
	\alpha^2  G(\tilde{\bq})u_y(\tilde{\bq})\int_{\tilde{\bf k}}
	\left(\frac{p_y}{p_x}\right)
	\left(\frac{h_y}{h_x}\right) G(\tilde{\bp}) C(\tilde{\bh})\,,\label{SM_Graph_b}
\end{equation}
from which we get the following contribution to $G(\tilde{\bq})q_x^2u_y^<(\tilde{\bq})$ in (\ref{SM_EOM_3}):
\begin{equation}
	\frac{\alpha^2  D}{8} G(\tilde{\bq}) q_x^2u_y(\tilde{\bq})\int{\dd^dk\over (2\pi)^d}
	\frac{\left(-\alpha^2 k_y^6-14 \alpha \mu k_x^4 k_y^4 +3\mu^2 k_x^8k_y^2 \right)}{\left(\alpha k_y^2+\mu k_x^4 \right)^4}=	\frac{\alpha^2  D}{8} G(\tilde{\bq}) q_x^2u_y(\tilde{\bq})(-I_3-14I_4+3I_5)\,.\label{SM_Graph_b1}
\end{equation}

Graph (c) in Fig. \ref{fig:propagator} represents
\begin{equation}
	\alpha^2  G(\tilde{\bq}) u_y(\tilde{\bq})\int_{\tilde{\bf k}}
	\left(\frac{h_y}{h_x}\right)^2G(\tilde{\bp}) C(\tilde{\bh})
\end{equation}
from which we get the following contribution to $G(\tilde{\bq})q_x^2u_y^<(\tilde{\bq})$ in (\ref{SM_EOM_3}):
\begin{equation}
	\frac{\alpha^2  D}{8} G(\tilde{\bq}) q_x^2u_y(\tilde{\bq})\int{\dd^dk\over (2\pi)^d}
	\frac{\left(-3\alpha^2 k_y^6-10 \alpha \mu k_x^4 k_y^4 +9\mu^2 k_x^8k_y^2 \right)}{\left(\alpha k_y^2+\mu k_x^4 \right)^4}=	\frac{\alpha^2  D}{8} G(\tilde{\bq}) q_x^2u_y(\tilde{\bq})(-3I_3-10I_4+9I_5)\,.\label{SM_Graph_c}
\end{equation}
Graph (d) in Fig. \ref{fig:propagator} represents
\begin{equation}
	\alpha^2  G(\tilde{\bq})u_y(\tilde{\bq})\int_{\tilde{\bf k}}
	\left(\frac{p_y}{p_x}\right)
	\left(\frac{h_y}{h_x}\right) G(\tilde{\bp}) C(\tilde{\bh})
\end{equation}
which is exactly the same as (\ref{SM_Graph_b}) for graph (b). Therefore, this graph also yields
\begin{equation}
	\frac{\alpha^2  D}{8} G(\tilde{\bq}) q_x^2u_y(\tilde{\bq})\int{\dd^dk\over (2\pi)^d}
	\frac{\left(-\alpha^2 k_y^6-14 \alpha \mu k_x^4 k_y^4 +3\mu^2 k_x^8k_y^2 \right)}{\left(\alpha k_y^2+\mu k_x^4 \right)^4}=	\frac{\alpha^2  D}{8} G(\tilde{\bq}) q_x^2u_y(\tilde{\bq})(-I_3-14I_4+3I_5)\,.\label{SM_Graph_d}
\end{equation}

Summing up (\ref{SM_Graph_a}), (\ref{SM_Graph_b1}), (\ref{SM_Graph_c}), and (\ref{SM_Graph_d}), we get the total contribution to $G(\tilde{\bq})q_x^2u_y^<(\tilde{\bq})$ in (\ref{SM_EOM_3}):
\beq
\frac{\alpha^2  D}{8} G(\tilde{\bq}) q_x^2u_y(\tilde{\bq})(-48I_4+16I_5)\,,
\eeq
which implies
\begin{equation}
	\delta\mu^{dir}=	-2{\alpha^2  D}(-3I_4+I_5)\,.\label{SM_mu}
\end{equation}

Using the results for $I_{4,5}$ calculated in Sec. \ref{Sec: Integrals} we get
\beqn
\mbox{Uncontrolled approximations with}~ k_x\in \{-\infty, \infty\}:~~~\delta \mu^{dir}=\mu g^u_\parallel \dd\ell\,,\\
\mbox{Uncontrolled approximations with}~ k_y\in \{-\infty, \infty\}:~~~\delta \mu^{dir}={1\over 2} \mu g^u_\perp \dd\ell\,,\\
\mbox{$\epsilon$ expansion with}~ k_x\in \{-\infty, \infty\}:~~~\delta \mu^{dir}= \mu g_h \dd\ell\,,
\eeqn
which implies
\beqn
\mbox{Uncontrolled approximations with}~ k_x\in \{-\infty, \infty\}:~~~\eta_\mu^{dir}=g^u_\parallel\,,\label{SM_D_Mu_Ux}\\
\mbox{Uncontrolled approximations with}~ k_y\in \{-\infty, \infty\}:~~~\eta_\mu^{dir}={1\over 2}g^u_\perp\,,\label{SM_D_Mu_Uy}\\
\mbox{$\epsilon$ expansion with}~ k_x\in \{-\infty, \infty\}:~~~\eta_\mu^{dir}= g_h\,.\label{SM_Mu_Dir_Epsilon}
\eeqn

Inserting (\ref{SM_D_Mu_Ux}), (\ref{SM_D_Mu_Uy}), and (\ref{SM_Mu_Dir_Epsilon}) into (\ref{SM_Mu_Cor_1}), respectively, we obtain
\beqn
\mbox{Uncontrolled approximations with}~ k_x\in \{-\infty, \infty\}:~~~\eta_\mu=\frac{5}{8}  g^u_\parallel\,,\\
\mbox{Uncontrolled approximations with}~ k_y\in \{-\infty, \infty\}:~~~\eta_\mu=\frac{1}{4}  g^u_\perp\,,\\
\mbox{$\epsilon$ expansion with}~ k_x\in \{-\infty, \infty\}:~~~\eta_\mu=\frac{5}{8} g_h\,.
\eeqn

\section{Calculation of integrals\label{Sec: Integrals}}
In this section we will calculate the integrals $I_{1,2,3,4,5}$   for the three different schemes.
\subsection{$I_1$}
This integral is
\begin{equation}
	I_1=\int{\dd^dk\over (2\pi)^d}
	\frac{k_y^2k_x^4}{\left( \mu k_x^4
		+\alpha k_y^2\right)^3}.
\end{equation}
\subsubsection{Uncontrolled $d=2$ with BZ $k_x\in\{-\infty,\infty\}$}
\begin{equation}
	I_1=\int_{\Lambda e^{-\zeta\dd\ell}}^\Lambda{\dd k_y\over 2\pi}\int_{-\infty}^\infty{\dd k_x\over 2\pi}
	\frac{k_y^2k_x^4}{\left( \mu k_x^4
		+\alpha k_y^2\right)^3}=\frac{3\zeta}{64\sqrt{2}\pi}\Lambda^{-1/2}\alpha^{-7/4}\mu^{-5/4}\dd\ell=\frac{3}{8}\frac{1}{\alpha^2 D}g^u_\parallel \dd\ell
\end{equation}
\subsubsection{Uncontrolled $d=2$ with BZ $k_y\in\{-\infty,\infty\}$}
\begin{equation}
	I_1=\int_{\Lambda e^{-\dd\ell}}^\Lambda{\dd k_x\over 2\pi}\int_{-\infty}^\infty{\dd k_y\over 2\pi}
	\frac{k_y^2k_x^4}{\left( \mu k_x^4
		+\alpha k_y^2\right)^3}=\frac{1}{16\pi}\Lambda^{-1}\alpha^{-3/2}\mu^{-3/2}\dd\ell=\frac{1}{4}\frac{1}{\alpha^2 D}g^u_\perp \dd\ell
\end{equation}
\subsubsection{Epsilon expansion with BZ $k_x\in\{-\infty,\infty\}$}
We extend the integral to $d$ dimensions by taking $y$ to be $d-1$ dimensional and $x$ 1 dimensional. Thus we have 
\begin{equation}
	I_1=	\int_{\Lambda e^{-\zeta\dd\ell}}^\Lambda{\dd^{d-1} k_y\over (2\pi)^{d-1}}\int_{-\infty}^\infty{\dd k_x\over 2\pi}
	\frac{k_y^2k_x^4}{\left( \mu k_x^4
		+\alpha k_y^2\right)^3}=\frac{3S_{d-1}\zeta}{64\sqrt{2}(2\pi)^{d-1}}\Lambda^{d-5/2}\alpha^{-7/4}\mu^{-5/4}d\ell=\frac{3}{8}\frac{1}{\alpha^2 D}g_h \dd\ell
\end{equation}

\subsection{$I_2$}
This integral is
\begin{equation}
	I_2=\int{\dd^dk\over (2\pi)^d}
	\frac{{k_x^4}}{(\mu k_x^4+\alpha k_y^2)^2}.
\end{equation}
\subsubsection{Uncontrolled $d=2$ with BZ $k_x\in\{-\infty,\infty\}$}
\begin{equation}
	I_2=\int_{\Lambda e^{-\zeta\dd\ell}}^\Lambda{\dd k_y\over 2\pi}\int_{-\infty}^\infty{\dd k_x\over 2\pi}
	\frac{{k_x^4}}{(\mu k_x^4+\alpha k_y^2)^2}=\frac{\zeta}{8\sqrt{2}\pi}\Lambda^{-1/2}\alpha^{-3/4}\mu^{-5/4}\dd\ell=\frac{1}{\alpha D}g^u_\parallel \dd\ell
\end{equation}
\subsubsection{Uncontrolled $d=2$ with BZ $k_y\in\{-\infty,\infty\}$}
\begin{equation}
	I_2=\int_{\Lambda e^{-\dd\ell}}^\Lambda{\dd k_x\over 2\pi}\int_{-\infty}^\infty{\dd k_y\over 2\pi}
	\frac{{k_x^4}}{(\mu k_x^4+\alpha k_y^2)^2}=\frac{1}{4\pi}\Lambda^{-1}\alpha^{-1/2}\mu^{-3/2}\dd\ell=\frac{1}{\alpha D}g^u_\perp \dd\ell
\end{equation}
\subsubsection{Epsilon expansion with BZ $k_x\in\{-\infty,\infty\}$}
\begin{equation}
	I_2=	\int_{\Lambda e^{-\zeta\dd\ell}}^\Lambda{\dd^{d-1} k_y\over (2\pi)^{d-1}}\int_{-\infty}^\infty{\dd k_x\over 2\pi}
	\frac{{k_x^4}}{(\mu k_x^4+\alpha k_y^2)^2}=\frac{S_{d-1}\zeta}{8\sqrt{2}(2\pi)^{d-1}}\Lambda^{d-5/2}\alpha^{-3/4}\mu^{-5/4}\dd\ell=\frac{1}{\alpha D}g_h \dd\ell
\end{equation}

\subsection{$I_3$}
This integral is
\begin{equation}
	I_3=\int{\dd^dk\over (2\pi)^d}
	\frac{ \alpha^2 k_y^6 }{\left(\alpha k_y^2+\mu k_x^4 \right)^4}.
\end{equation}
\subsubsection{Uncontrolled $d=2$ with BZ $k_x\in\{-\infty,\infty\}$}
\begin{equation}
	I_3=\int_{\Lambda e^{-\zeta\dd\ell}}^\Lambda{\dd k_y\over 2\pi}\int_{-\infty}^\infty{\dd k_x\over 2\pi}
	\frac{ \alpha^2 k_y^6 }{\left(\alpha k_y^2+\mu k_x^4 \right)^4}=\frac{77\zeta}{256\sqrt{2}\pi}\Lambda^{-1/2}\alpha^{-7/4}\mu^{-1/4}\dd\ell=\frac{77}{32}\frac{\mu}{\alpha^2 D}g^u_\parallel \dd\ell
\end{equation}
\subsubsection{Uncontrolled $d=2$ with BZ $k_y\in\{-\infty,\infty\}$}
\begin{equation}
	I_3=\int_{\Lambda e^{-\dd\ell}}^\Lambda{\dd k_x\over 2\pi}\int_{-\infty}^\infty{\dd k_y\over 2\pi}
	\frac{ \alpha^2 k_y^6 }{\left(\alpha k_y^2+\mu k_x^4 \right)^4}=\frac{5}{32\pi}\Lambda^{-1}\alpha^{-3/2}\mu^{-1/2}\dd\ell=\frac{5}{8}\frac{\mu}{\alpha^2 D}g^u_\perp \dd\ell
\end{equation}
\subsubsection{Epsilon expansion with BZ $k_x\in\{-\infty,\infty\}$}
\begin{equation}
	I_3=	\int_{\Lambda e^{-\zeta\dd\ell}}^\Lambda{\dd^{d-1} k_y\over (2\pi)^{d-1}}\int_{-\infty}^\infty{\dd k_x\over 2\pi}
	\frac{ \alpha^2 k_y^6 }{\left(\alpha k_y^2+\mu k_x^4 \right)^4}=\frac{77 S_{d-1}\zeta}{256\sqrt{2}(2\pi)^{d-1}}\Lambda^{d-5/2}\alpha^{-7/4}\mu^{-1/4}\dd\ell=\frac{77}{32}\frac{\mu}{\alpha^2 D}g_h \dd\ell
\end{equation}

\subsection{$I_4$}
This integral is
\begin{equation}
	I_4=\int{\dd^dk\over (2\pi)^d}
	\frac{\alpha \mu k_x^4 k_y^4 }{\left(\alpha k_y^2+\mu k_x^4 \right)^4}.
\end{equation}
\subsubsection{Uncontrolled $d=2$ with BZ $k_x\in\{-\infty,\infty\}$}
\begin{equation}
	I_4=\int_{\Lambda e^{-\zeta\dd\ell}}^\Lambda{\dd k_y\over 2\pi}\int_{-\infty}^\infty{\dd k_x\over 2\pi}
	\frac{\alpha \mu k_x^4 k_y^4 }{\left(\alpha k_y^2+\mu k_x^4 \right)^4}=\frac{7\zeta}{256\sqrt{2}\pi}\Lambda^{-1/2}\alpha^{-7/4}\mu^{-1/4}\dd\ell=\frac{7}{32}\frac{\mu}{\alpha^2 D}g^u_\parallel \dd\ell
\end{equation}
\subsubsection{Uncontrolled $d=2$ with BZ $k_y\in\{-\infty,\infty\}$}
\begin{equation}
	I_4=\int_{\Lambda e^{-\dd\ell}}^\Lambda{\dd k_x\over 2\pi}\int_{-\infty}^\infty{\dd k_y\over 2\pi}
	\frac{\alpha \mu k_x^4 k_y^4 }{\left(\alpha k_y^2+\mu k_x^4 \right)^4}=\frac{1}{32\pi}\Lambda^{-1}\alpha^{-3/2}\mu^{-1/2}\dd\ell=\frac{1}{8}\frac{\mu}{\alpha^2 D}g^u_\perp \dd\ell
\end{equation}
\subsubsection{Epsilon expansion with BZ $k_x\in\{-\infty,\infty\}$}
\begin{equation}
	I_4=	\int_{\Lambda e^{-\zeta\dd\ell}}^\Lambda{\dd^{d-1} k_y\over (2\pi)^{d-1}}\int_{-\infty}^\infty{\dd k_x\over 2\pi}
	\frac{\alpha \mu k_x^4 k_y^4 }{\left(\alpha k_y^2+\mu k_x^4 \right)^4}=\frac{7 S_{d-1}\zeta}{256\sqrt{2}(2\pi)^{d-1}}\Lambda^{d-5/2}\alpha^{-7/4}\mu^{-1/4}\dd\ell=\frac{7}{32}\frac{\mu}{\alpha^2 D}g_h \dd\ell
\end{equation}

\subsection{$I_5$}
This integral is
\begin{equation}
	I_5=\int{\dd^dk\over (2\pi)^d}
	\frac{\mu^2 k_x^8k_y^2}{\left(\alpha k_y^2+\mu k_x^4 \right)^4}.
\end{equation}
\subsubsection{Uncontrolled $d=2$ with BZ $k_x\in\{-\infty,\infty\}$}
\begin{equation}
	I_5=\int_{\Lambda e^{-\zeta\dd\ell}}^\Lambda{\dd k_y\over 2\pi}\int_{-\infty}^\infty{\dd k_x\over 2\pi}
	\frac{\mu^2 k_x^8k_y^2}{\left(\alpha k_y^2+\mu k_x^4 \right)^4}=\frac{5\zeta}{256\sqrt{2}\pi}\Lambda^{-1/2}\alpha^{-7/4}\mu^{-1/4}\dd\ell=\frac{5}{32}\frac{\mu}{\alpha^2 D}g^u_\parallel \dd\ell
\end{equation}
\subsubsection{Uncontrolled $d=2$ with BZ $k_y\in\{-\infty,\infty\}$}
\begin{equation}
	I_5=\int_{\Lambda e^{-\dd\ell}}^\Lambda{\dd k_x\over 2\pi}\int_{-\infty}^\infty{\dd k_y\over 2\pi}
	\frac{\mu^2 k_x^8k_y^2}{\left(\alpha k_y^2+\mu k_x^4 \right)^4}=\frac{1}{32\pi}\Lambda^{-1}\alpha^{-3/2}\mu^{-1/2}\dd\ell=\frac{1}{8}\frac{\mu}{\alpha^2 D}g^u_\perp \dd\ell
\end{equation}
\subsubsection{Epsilon expansion with BZ $k_x\in\{-\infty,\infty\}$}
\begin{equation}
	I_5=	\int_{\Lambda e^{-\zeta\dd\ell}}^\Lambda{\dd^{d-1} k_y\over (2\pi)^{d-1}}\int_{-\infty}^\infty{\dd k_x\over 2\pi}
	\frac{\mu^2 k_x^8k_y^d}{\left(\alpha k_y^2+\mu k_x^4 \right)^4}=\frac{5 S_{d-1}\zeta}{256\sqrt{2}(2\pi)^{d-1}}\Lambda^{d-5/2}\alpha^{-7/4}\mu^{-1/4}\dd\ell=\frac{5}{32}\frac{\mu}{\alpha^2 D}g_h\dd\ell
\end{equation}


\begin{thebibliography}{16}%
\makeatletter
\providecommand \@ifxundefined [1]{%
 \@ifx{#1\undefined}
}%
\providecommand \@ifnum [1]{%
 \ifnum #1\expandafter \@firstoftwo
 \else \expandafter \@secondoftwo
 \fi
}%
\providecommand \@ifx [1]{%
 \ifx #1\expandafter \@firstoftwo
 \else \expandafter \@secondoftwo
 \fi
}%
\providecommand \natexlab [1]{#1}%
\providecommand \enquote  [1]{``#1''}%
\providecommand \bibnamefont  [1]{#1}%
\providecommand \bibfnamefont [1]{#1}%
\providecommand \citenamefont [1]{#1}%
\providecommand \href@noop [0]{\@secondoftwo}%
\providecommand \href [0]{\begingroup \@sanitize@url \@href}%
\providecommand \@href[1]{\@@startlink{#1}\@@href}%
\providecommand \@@href[1]{\endgroup#1\@@endlink}%
\providecommand \@sanitize@url [0]{\catcode `\\12\catcode `\$12\catcode
  `\&12\catcode `\#12\catcode `\^12\catcode `\_12\catcode `\%12\relax}%
\providecommand \@@startlink[1]{}%
\providecommand \@@endlink[0]{}%
\providecommand \url  [0]{\begingroup\@sanitize@url \@url }%
\providecommand \@url [1]{\endgroup\@href {#1}{\urlprefix }}%
\providecommand \urlprefix  [0]{URL }%
\providecommand \Eprint [0]{\href }%
\providecommand \doibase [0]{http://dx.doi.org/}%
\providecommand \selectlanguage [0]{\@gobble}%
\providecommand \bibinfo  [0]{\@secondoftwo}%
\providecommand \bibfield  [0]{\@secondoftwo}%
\providecommand \translation [1]{[#1]}%
\providecommand \BibitemOpen [0]{}%
\providecommand \bibitemStop [0]{}%
\providecommand \bibitemNoStop [0]{.\EOS\space}%
\providecommand \EOS [0]{\spacefactor3000\relax}%
\providecommand \BibitemShut  [1]{\csname bibitem#1\endcsname}%
\let\auto@bib@innerbib\@empty

\bibitem{MW}
N. D. Mermin and H. Wagner,
``Absence of ferromagnetism or antiferromagnetism in one- or two-dimensional isotropic Heisenberg Models,'' Physical Review Letters {\bf 17,} 1133 (1966); P.C. Hohenberg, ``Existence of long-range order in one and two dimensions,'' Physical Review {\bf 158,} 383 (1967).
 \bibitem [{\citenamefont {Vicsek}\ \emph {et~al.}(1995)\citenamefont {Vicsek},
  \citenamefont {Czir{\'{o}}k}, \citenamefont {Ben-Jacob}, \citenamefont
  {Cohen},\ and\ \citenamefont {Shochet}}]{vicsek_prl95}%
  \BibitemOpen
  \bibfield  {author} {\bibinfo {author} {\bibfnamefont {T.}\
  \bibnamefont {Vicsek}}, \bibinfo {author} {\bibfnamefont {A.}\
  \bibnamefont {Czir{\'{o}}k}}, \bibinfo {author} {\bibfnamefont {E.}\
  \bibnamefont {Ben-Jacob}}, \bibinfo {author} {\bibfnamefont {I.}\
  \bibnamefont {Cohen}}, \ and\ \bibinfo {author} {\bibfnamefont {O.}\
  \bibnamefont {Shochet}},\ }\bibfield  {title} {\enquote {\bibinfo {title}
  {{Novel Type of Phase Transition in a System of Self-Driven Particles}},}\
  }\href {\doibase citeulike-article-id:1648582 doi:
  10.1103/PhysRevLett.75.1226} {\bibfield  {journal} {\bibinfo  {journal}
  {Physical Review Letters}\ }\textbf {\bibinfo {volume} {75}},\ \bibinfo
  {pages} {1226--1229} (\bibinfo {year} {1995})}\BibitemShut {NoStop}%
   \bibitem [{\citenamefont {Toner}\ and\ \citenamefont {Tu}(1995)}]{toner_prl95}%
  \BibitemOpen
  \bibfield  {author} {\bibinfo {author} {\bibfnamefont {J.}\ \bibnamefont
  {Toner}}\ and\ \bibinfo {author} {\bibfnamefont {Y.}\ \bibnamefont {Tu}},\
  }\bibfield  {title} {\enquote {\bibinfo {title} {{Long-Range Order in a
  Two-Dimensional Dynamical $XY$ Model: How Birds Fly Together}},}\ }\href
  {\doibase 10.1103/PhysRevLett.75.4326} {\bibfield  {journal} {\bibinfo
  {journal} {Physical Review Letters}\ }\textbf {\bibinfo {volume} {75}},\
  \bibinfo {pages} {4326--4329} (\bibinfo {year} {1995})}\BibitemShut {NoStop}%
\bibitem [{\citenamefont {Toner}\ and\ \citenamefont {Tu}(1998)}]{toner_pre98}%
  \BibitemOpen
  \bibfield  {author} {\bibinfo {author} {\bibfnamefont {J.}\ \bibnamefont
  {Toner}}\ and\ \bibinfo {author} {\bibfnamefont {Y.}\ \bibnamefont {Tu}},\
  }\bibfield  {title} {\enquote {\bibinfo {title} {{Flocks, herds, and schools:
  A quantitative theory of flocking}},}\ }\href {\doibase
  10.1103/PhysRevE.58.4828} {\bibfield  {journal} {\bibinfo  {journal}
  {Physical Review E}\ }\textbf {\bibinfo {volume} {58}},\ \bibinfo {pages}
  {4828--4858} (\bibinfo {year} {1998})}\BibitemShut {NoStop}%
  \bibitem{toner_pre12}
J.~Toner,  ``Reanalysis of the hydrodynamic theory of fluid, polar-ordered flocks,"
  Physical Review E {\bf 86}, 031918 (1012).

\bibitem{Chate_DADAM_rev}
H.~Chat\'{e}, ``Dry Aligning Dilute Active Matter," Annual Review of Condensed Matter Physics {\bf 11}, 189 (2020).


 \bibitem{inc_trans} L. Chen,  J. Toner,  and C. F. Lee,  ``Critical phenomenon of the order-disorder transition in incompressible flocks,''  New Journal of Physics {\bf 17},  042002 (2015).



   \bibitem [{\citenamefont {Chen}\ \emph {et~al.}(2016)\citenamefont {Chen},
  \citenamefont {Lee},\ and\ \citenamefont {Toner}}]{CLT_Ncomm}%
  \BibitemOpen
  \bibfield  {author} {\bibinfo {author} {\bibfnamefont {L.}\ \bibnamefont
  {Chen}}, \bibinfo {author} {\bibfnamefont {C.~F.}\ \bibnamefont {Lee}}, \
  and\ \bibinfo {author} {\bibfnamefont {J.}\ \bibnamefont {Toner}},\
  }\bibfield  {title} {\enquote {\bibinfo {title} {{Mapping two-dimensional
  polar active fluids to two-dimensional soap and one-dimensional
  sandblasting}},}\ }\href {\doibase 10.1038/ncomms12215} {\bibfield  {journal}
  {\bibinfo  {journal} {Nature Communications}\ }\textbf {\bibinfo {volume}
  {7}},\ \bibinfo {pages} {12215} (\bibinfo {year} {2016.})}



  \bibitem{Chen} L. Chen, C. F. Lee, and J. Toner, ``Incompressible polar active
  fluids in the moving phase in dimensions $d > 2$," New
  Journal of Physics {\bf 20},
  113035 (2018).

        \bibitem{Ano_pol} A. Maitra, P. Srivastava, M. C. Marchetti, S. Ramaswamy, M. Lenz, ``Swimmer suspensions on substrates: anomalous stability and long-range order,'' Physical Review Letters {\bf 124}, 028002 (2020).




\bibitem{CLMT-PRL1}
L. Chen, C. F. Lee, A. Maitra, and J. Toner, ``Packed Swarms on Dirt: Two-Dimensional Incompressible Flocks with Quenched and Annealed Disorder,''  Physical Review Letters {\bf 129}, 188004 (2022).

\bibitem{CLMT-PRE1}
L. Chen, C. F. Lee, A. Maitra, and J. Toner, ``Hydrodynamic theory of two-dimensional incompressible polar active fluids with quenched and annealed disorder, " Physical Review E {\bf 106}, 044608 (2022).

\bibitem{Toner_Malthus}
J.~Toner, ``Birth, Death, and Flight: A Theory of Malthusian Flocks,"
Physical Review Letters {\bf 108}, 088102 (2012).

\bibitem [{\citenamefont {Chen}\ \emph
  {et~al.}(2020{\natexlab{a}})\citenamefont {Chen}, \citenamefont {Lee},\ and\
  \citenamefont {Toner}}]{CLT_malthus1}%
  \BibitemOpen
  \bibfield  {author} {\bibinfo {author} {\bibfnamefont {L.}\ \bibnamefont
  {Chen}}, \bibinfo {author} {\bibfnamefont {C.~F.}\ \bibnamefont {Lee}}, \
  and\ \bibinfo {author} {\bibfnamefont {J.}\ \bibnamefont {Toner}},\
  }\bibfield  {title} {\enquote {\bibinfo {title} {{Moving, Reproducing, and
  Dying Beyond Flatland: Malthusian Flocks in Dimensions $d > 2$}},}\ }\href
  {\doibase 10.1103/PhysRevLett.125.098003} {\bibfield  {journal} {\bibinfo
  {journal} {Physical Review Letters}\ }\textbf {\bibinfo {volume} {125}},\
  \bibinfo {pages} {098003} (\bibinfo {year} {2020}{\natexlab{a}})},\ \Eprint
  {http://arxiv.org/abs/2001.01300} {arXiv:2001.01300} \BibitemShut {NoStop}%
\bibitem [{\citenamefont {Chen}\ \emph
  {et~al.}(2020{\natexlab{b}})\citenamefont {Chen}, \citenamefont {Lee},\ and\
  \citenamefont {Toner}}]{CLT_malthus2}%
  \BibitemOpen
  \bibfield  {author} {\bibinfo {author} {\bibfnamefont {L.}\ \bibnamefont
  {Chen}}, \bibinfo {author} {\bibfnamefont {C.~F.}\ \bibnamefont {Lee}}, \
  and\ \bibinfo {author} {\bibfnamefont {J.}\ \bibnamefont {Toner}},\
  }\bibfield  {title} {\enquote {\bibinfo {title} {{Universality class for a
  nonequilibrium state of matter: A $d=4-\epsilon$ expansion study
  of Malthusian flocks}},}\ }\href {\doibase 10.1103/PhysRevE.102.022610}
  {\bibfield  {journal} {\bibinfo  {journal} {Physical Review E}\ }\textbf
  {\bibinfo {volume} {102}},\ \bibinfo {pages} {022610} (\bibinfo {year}
  {2020}{\natexlab{b}})}.







%


\bibitem{Kashuba} A. Kashuba, ``Exact scaling of spin-wave correlations in the 2D XY ferromagnet with dipolar forces,''  Physical Review Letters {\bf 73}, 2264 (1994).



\bibitem{Golub}
L. Golubovi\'{c} and Z.-G. Wang, ``Anharmonic elasticity of smectics A and the Kardar-Parisi-Zhang model,"
Physical Review Letters {\bf 69,} 2535 (1992);
``Kardar-Parisi-Zhang model and anomalous elasticity of two- and three-dimensional smectic-$A$ liquid crystals," Physical Review E {\bf 49,} 2567 (1994).

\bibitem{KPZ}
M. Kardar, G. Parisi and Y.-C. Zhang, ``Dynamic scaling of growing interfaces,"
Physical Review Letters {\bf 56}, 889-892 (1986);

\bibitem{foot1}
Alternatatively, we could act on eqn. (\ref{eq:u_eom}) with $P_{xi}(\bq)$ to get the EOM for $u_x(\tilde{\bq})$. However, the resultant equation is equivalent to (\ref{2Dy4}), since $u_x(\tilde{\bq})$ and $u_y(\tilde{\bq})$ are not independent,  but rather related by the incompressibility constraint $q_xu_x(\tilde{\bq})+q_yu_y(\tilde{\bq})=0$.

 \bibitem [{\citenamefont {Forster}\ \emph {et~al.}(1977)\citenamefont
  {Forster}, \citenamefont {Nelson},\ and\ \citenamefont
  {Stephen}}]{forster_pra77}
  \BibitemOpen
  \bibfield  {author} {\bibinfo {author} {\bibfnamefont {D.}\ \bibnamefont
  {Forster}}, \bibinfo {author} {\bibfnamefont {D.~R.}\ \bibnamefont
  {Nelson}}, \ and\ \bibinfo {author} {\bibfnamefont {M.~J.}\ \bibnamefont
  {Stephen}},\ }\bibfield  {title} {\enquote {\bibinfo {title} {{Large-distance
  and long-time properties of a randomly stirred fluid}},}\ }\href {\doibase
  10.1103/PhysRevA.16.732} {\bibfield  {journal} {\bibinfo  {journal} {Physical
  Review A}\ }\textbf {\bibinfo {volume} {16}},\ \bibinfo {pages} {732--749}
  (\bibinfo {year} {1977})}\BibitemShut {NoStop}%
\bibitem{PCMP} P. M. Chaikin and T. C. Lubensky, {\it Principles of Condensed Matter Physics}, (Cambridge University Press, Cambridge, UK, 1995).

\bibitem{Hohenberg} P. C. Hohenberg and B. I. Halperin, ``Theory of dynamic ctitical phenomena,'' Review of Modern Physics {\bf 49}, 435 (1977).

\bibitem{Nelson_traj} D. R. Nelson, ``Crossover scaling functions and renormalization-group trajectory integrals,'' Physical Review B{\bf 11}, 3504 (1975).

%
%
%
%
%
%
%
%
%
%
%
%
%



\end{thebibliography}

\begin{thebibliography}{99}
								
								
								\bibitem{forster_pra77}
								D. Forster, D. R. Nelson, and M. J. Stephen, ``Large-
								distance and long-time properties of a randomly stirred
								fluid'', Physical Review A {\bf 16}, 732-749 (1977).
								
							\end{thebibliography}
\end{document}